# GUIDE
for a blended learning system

Bachelor

Masters

Doctorate

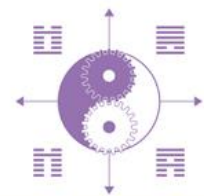

La gouvernance en mouvement

AGENCE UNIVERSITAIRE DE LA FRANCOPHONIE

*Longer journeys start with a first step.*

Proverb

Elaborated by:

Mokhtar Ben Henda

Published by:

Agence universitaire de la Francophonie, Direction régionale Asie-Pacifique (AUF / DRAP)

Terms of use:

This guide is licensed under the Creative Commons CC-BY-SA (Attribution - Sharing under the Same Conditions) 4.0 International

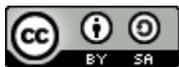

Edition: June 2020

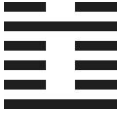 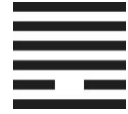

**Initial difficulties:** uncertainties inherent in a beginning period. It's about overcoming hesitation and indecision without falling into haste. Taking the time to reflect in the service of a strategy remains the fairest path.

**Getting along with everyone:** dialogue with others in a constructive approach. It is a question of making its specificity understood by welcoming that of the other, and thus making it possible to register these differences in a harmonious whole, because they converge. Get along despite and from these specificities.

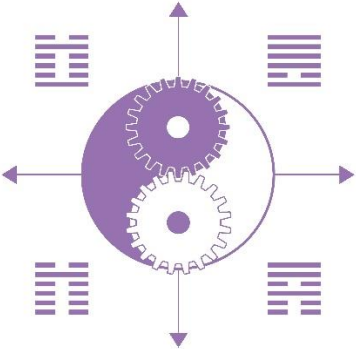

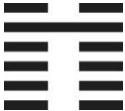 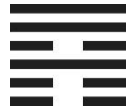

**Alliance:** harmony of things from a discordant whole. It is about identifying a center around which disparate elements converge. It is from different although complementary components that this convergent center is organized, as it is animated by a vision that goes beyond them.

**Progress step by step:** past achievements serving the accomplishment of present actions. It is a question of measuring the benefits of the efforts made by relying on the passage of time. Build on what has been achieved in order to define the forces from which the outcome of a strategy becomes possible.

The book of changes

# FOREWORD

This guide is proposed as an operational instrument for CONFRASIE member universities (Regional Rectors' Conference of AUF member institutions in Pacific-Asia) in their projects to set up a blended learning system for Bachelor's, Master's or Doctorate degrees. It is structured in sections corresponding to a complete process of operationalizing a blended learning system, from the definition of an implementation strategy to the assessment of results.

This guide covers also conceptual and theoretical fundamentals of distance learning as well as methodological and procedural tips and recommendations on how to implement blended learning in an existing face-to-face curriculum. It can serve for leaders of educational ICT-based projects as a guidance document to take pedagogical, technological and methodological decisions for the development, monitoring and assessment of a blended learning curricula. This guide can be augmented by other standards, tool and software manuals offering further training materials and guidelines on educational skills and services.

It is therefore up to the institution to consider the application of this Guide, to define a quality approach in favor of the planned blended learning project: objectives, expected and measurable outcomes, resources necessary to achieve concrete fallouts, mechanisms for assessment and validation of results. This supposes the definition of the main strategic trends of the blended learning project and their integration in the governance model of the institution (development plan* / action plan*). It is strongly recommended to start with the definition of a legal and regulatory basis for the project before thinking about a business model, in conformity with the national regulations, that suits the institution potentials and the service providers economic expectations. Any institutionalization of the blended learning system would depend of both prerequisites.

It is also up to the institution to ensure the effectiveness, efficacy and efficiency of the corrective and innovative measures to implement before, during and after the implementation of the project. This implies integrating, as soon as the blended learning offer is put in place, a cyclic system for evaluating its piloting and outcomes.

This guide was developed by Mr. Mokhtar BEN HENDA, teacher-researcher at the University of Bordeaux Montaigne (France). During the writing of this guide, several contributors from the member institutions of CONFRASIE cooperated at several stages of its production, drawing up directives, providing advice, carrying out content and form revisions and participating in feasibility tests.

# SUMMARY

# TABLE OF ILLUSTRATIONS

# ABBREVIATIONS AND ACRONYMS

| | |
|---|---|
| AUF | Agence Universitaire de la Francophonie |
| CMS | Content Management System |
| CNF | Campus Numérique Francophone |
| CNFp | Campus Numérique Francophone partner |
| CONFRASIE | Regional Rectors' Conference of AUF member institutions in Pacific-Asia |
| CPU | Central Processing Unit |
| DRAP | AUF Regional Directorate in Pacific-Asia |
| FAQ | Frequently Asked Questions |
| FOAD | Formation Ouverte et A Distance (Open Distance Training) |
| LCMS | Learning Content Management System |
| LMS | Learning Management System |
| MOOC | Massive Open Online Course |
| SCORM | Sharable Content Object Reference Model |
| ICTE | Information and Communication Technologies for Education |
| URL | Uniform Resource Locator |
| VPS | Custom Virtual Server |

# GENERAL FRAMEWORK

The international context of education is undergoing a profound change, brought about by the emergence of information and communication technologies (ICT) and the use of constructivist pedagogical methods found in ICT and distance education (e -Learning) a factor of support and promotion on a large scale. Distance education, as a form of knowledge dissemination, formerly practiced via postal correspondence or educational television and radio, is now regenerating as a result of ICT use, in new forms of space and time abstraction and virtualization.

With digital technologies, we can now identify two main models of distance education: bimodal and mixed (also called blended or hybrid). The bimodal one refers to institutions which, alongside their traditional face-to-face teaching, offer separate distance learning, often for a different audience, using ICT and e-learning widely.

The mixed or blended model seems to become the dominant choice of contemporary approaches to teach or train using ICT. It does not entail the creation of two distinct entities within educational institutions but rather consists of "a formal education program that's made up of in-person classroom time as well as individual study online using eLearning software. It is a type of multichannel method that incorporates tutor*-led activities, images, video, digital tasks and face-to-face discussion."[1].

Blended teaching or training thus allows any teaching or training structure to attempt the integration of digital technologies into its educational practices by associating with existing face-to-face teaching, a part of online (distance) teaching. This association, or assembly, carried out in various proportions (cf. Figure 1), generally results from several factors. In addition to the pedagogical and didactic interests attributed to ICT and ICTE (ICT for Education), there is notably a need to control costs, better manage learning spaces and times, introduce innovative teaching methods and training to better prepare students for the demands of the job market and socio-professional integration.

The transition from a traditional pedagogical model to a blended one, however, requires a set of measures and institutional conditions which can sometimes be burdensome from a legal, economic, technological and pedagogical point of view.

## 1 LEGAL FRAMEWORK OF BLENDED LEARNING

It is still common for the use of ICT in pedagogical teaching and training practices to be an individual initiative of teachers and trainers who apply their own social ICT practices into their teachings. Contacts with students by email, dissemination of course content by blogs or on the cloud, discussions via forums and social networks, etc. are always alternatives to blend courses without these procedures being recognized or validated by internal regulatory texts and even less by national legal context.

---

[1] Team, C. A. E. (2016). What is Blended Learning Method and how it works? CAE Computer Aided E-Learning. https://www.cae.net/blended-learning-introduction/

Even today, a very large number of institutions in many countries of the world - including in Southeast Asia - have not yet taken the step to recognize ICT and distant activities as an integral part of the learning offer. A legal vacuum still blocks the outcome of several online learning projects and prevents them from recognizing the value of the diplomas obtained by distance education.

In the South-East Asia region, Vietnam presents positive indicators which are gradually moving in the direction of promoting online learning. An example would be the circular[2] that the "Ministry of Education and Training adopted on April 22, 2016… relating to the application of information technology in the management and organization of online learning »[3] or the Ministry Distance Training Regulations dated 28/4/2017 in which blended learning is mentioned[4]. These texts will undoubtedly evolve towards stronger legal measures such as a decree (or a law) which would take into account the strategic nature of online learning in the national socio-economic landscape. This was the case of France which, by decree n ° 2017-619[5] of 24 April 2017 on the provision of distance education in higher education institutions, changed the previous texts towards recognition of online learning as a national educational alternative in higher education institutions in France (see box below).

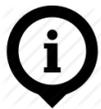

"Art. OC 611-10. - The lessons delivered within the framework of training in higher education institutions can be provided either in the presence of users, or remotely, if necessary, in digital form, or according to systems combining the two forms.

"A minimum volume of pedagogical education, set by regulation, can be provided in the presence of students.

"Art. OC 611-11. - Constitutes a teaching of higher education by distance teaching delivered outside the physical presence in the same place as the student of the teacher who dispenses it. This education is wholly or mainly designed and organized by teachers from the institution that offers it.

"Distance education is accompanied by personalized support for students.

"Art. OC 611-12. - The conditions for the validation of lessons, given in the presence of users or remotely, if necessary, in digital form, are fixed in each higher education institution at the latest at the end of the first month of the teaching year and they cannot be changed during the year.

"The validation of the lessons controlled by tests organized remotely in digital form, must be guaranteed by:

"1 ° Verification that the candidate has the technical means enabling him to pass the tests effectively;

"2 ° Verification of the identity of the candidate;

"3 ° Monitoring the event and respecting the rules applicable to exams. "

---

[2] Bộ Giáo dục và Đào tạo. (2016). Thông tư số 12/2016/TT-BGDĐT Quy định Ứng dụng công nghệ thông tin trong quản lý, tổ chức đào tạo qua mạng
[3] NGUYEN TAN Dai (2017), « Les TIC au service de la qualité des formations : le cas des programmes vietnamiens évalués par l'ASEAN University Network ». Thèse de doctorat, Université de Strasbourg
[4] Quy chế về ĐT từ xa Ban hành kèm theo Thông tư số 10/2017/TT-BGDĐT ngày 28 tháng 4 năm 2017 của Bộ trưởng Bộ Giáo dục và Đào tạo
[5] NOR:MENS1707714D - JORF n°0098 du 26 avril 2017, texte n° 10 - Décret n° 2017-619 du 24 avril 2017 relatif à la mise à disposition d'enseignements à distance dans les établissements d'enseignement supérieur. https://www.legifrance.gouv.fr/eli/decret/2017/4/24/MENS1707714D/jo/texte

The text of this decree allows a margin of freedom for university institutions to define internally the conditions for the validation of distance learning. In the absence of a decree of national scope, and according to known practices in the event of a legal vacuum, the institutions can appeal to an internal regulation in favor of the online learning without contradicting the legislation in force. This can take the form of an internal note, a recommendation from the scientific council, or an additional clause in the institution's action or strategic development plans*.

## 2 BUSINESS MODELS OF BLENDED LEARNING

The second stumbling block for online learning is the business model to be put in place to justify the expenses incurred around distance education activities. This is what Arnaud Coulon and Michel Ravailhe call "Training system economy"[6] which regularly appears as one of the most complex tasks in the setting up of a blended learning, because the calculation of the costs of online learning is facing several pitfalls and depends on several variables that are difficult to model and measure. In addition, when we try to identify "roughly" these costs, it is very difficult to model their structure because the online learning systems are unique and their financial arrangements too.

### 2.1 PRIVAT/PUBLIC, FOR-PROFIT/NON-PROFIT UNIVERSITIES

One major difference is essentially observed among for-profit and non-profit institutions as indicated in a report produced in 2013 by Hanover Research Center[7] in which it examines the business models that for-profit, non-profit, and "open-source" higher education institutions are currently implementing for online education.

The primary difference between them is described as follows: "While traditional colleges and universities rely heavily on government appropriations and private donations, for-profits must be self-sufficient and respond to market forces to be successful. The market-place naturally forces for-profit institutions to offer an educational product that is valuable to students and to do so at a reasonable price. Traditional institutions, however, are not always subject to this threat of "creative destruction". The recent growth and success of for-profits at a time when many traditional universities are struggling financially serves as a testament to the viability of the sector"[8].

In order to compensate for the loss of operating funds, either due to budget cuts (in the case of public institutions) or reductions in the size of an endowment (for private institutions), some universities have decided to enter the online education world by extending courses or entire degrees into the online market. For more precision, we recommend to read the Hanover Research center report cited in footnote.

Nevertheless, in any situation, common constraints and pitfalls apply for accurate estimation of cost effects of an eLearning system.

---

[6] Coulon A. & Ravailhe M. (2003). « FOAD : économie des dispositifs et calcul des coûts ». https://www.centre-inffo.fr/IMG/pdf/economie_et_calcul_des_couts_foad.pdf
[7] Hanover Research Center (2013). "Business Models for Online Higher Education".
https://www.hanoverresearch.com/media/Business-Models-for-Online-Higher-Education-1.pdf
[8] Bennett D. et.al. (2010). "For-Profit Higher Education: Growth, Innovation and Regulation". Center for College Affordability and Productivity. http://www.centerforcollegeaffordability.org/uploads/ForProfit_HigherEd.pdf

## 2.2 COMMON CONSTRAINTS FOR ONLINE LEARNING COST EFFECT

In the university environment, several constraints characterize the situation with regard to the question of the online learning costs in a blended learning system. We quote some of them:

- The missing of an accounting model (like that of the face-to-face teaching) to remunerate online activities;
- The distance learning time and the control of its effective duration (as a time deemed necessary) poses the problem of the scale to be used for counting one hour of work;
- Evidence of the training action which no longer obeys observable and quantifiable measures of physical presence in the classroom;
- The new functions of remote tutors* / trainers are not recognized in the models of face-to-face learning;
- The specific remote monitoring methods (messaging, forum, virtual class, chat, etc.) are not commonly counted as virtual workload in the teachers' regulatory department …

In fact, the system of values and the logic of public funding in which the university is part, as well as the methods of allocating resources through the use of standards whose rationality appears essentially administrative, do not allow tackle the issue of costs very easily, due to the difficulty of measuring the reality of expenditure;

- In universities, particularly in the public sector, we very often come up against problems of data collection and identification of the real costs linked to online learning. A margin of opacity on costs is linked to many factors such as the weakness of the existing management instrumentation, local management practices which are sometimes "artisanal"; the breakdown of charges which can be random and arbitrary; charging expenditure to different budgets depending on different services, underestimating and undervaluing the contribution of teachers - researchers, etc.
- In several contexts, including in countries with advanced technological traditions, the time spent teaching at distance through technology is not recognized as a measurable activity in the same way as face-to-face teaching that can be calculated (and paid for) in number of hours spent in class with students.

The main reason for the complexity of the online learning business model is often the legal vacuum around the distance education activity that prevents the university or training organization from managing the real costs of an online learning. In this case, several institutions proceed by consensus (circular agreement, internal regulations, memorandum, etc.), or even by case law, to define a regulatory framework allowing accounting management of the online learning.

Several economic models of online learning are however proposed, including the cost-per-activity method (or ABC: Activity Based Costing). Initiated in the business world then brought back in the context of online learning, the ABC method has advantages as long as it allows to find a solution to the cost problem, but also has disadvantages as long as it is originally based on an entrepreneurial reasoning on return on investment (ROI). However, applied to online learning, the ABC method can show that it is not the products which constitute the main cost, but rather the activities which are generally subdivided into three categories:

1. Production activities that directly contribute to the realization of the online learning product / service. This may include the costs of producing the teaching material (courses and multimedia supports for teaching resources; Quizzes, teaching activities, course scenarios, assessment material for tutoring*, etc.);
2. Support activities that help production activities to fulfill their role. This may include, for example, the cost of logistics used such as IT, secretarial work, as well as management and maintenance costs such as verification and validation of form and content, all calculated in terms of hours or days / men according to the cost scales in use;
3. Structural activities that contribute to determining objectives, defining and implementing training and monitoring its performance. This can be assessed in terms of cost / time of the management team and the quality management team, etc.

To properly identify the costs of an online learning through this model, we must first develop a complete mapping of learning activities and their respective tasks in order to identify the costs. The learning manager must imperatively participate in the institution of the list of activities selected after possible simplifications. In this case, questionnaires and interviews play an interesting role in modeling the structure of activities.

The following grid, developed as part of an analysis of the costs of an online learning at Jules Verne University of Picardie in France[9], offers a set of categories of activities deemed to consume specific and shared costs.

| Activities | Specific charges | Common charges | | Total |
|---|---|---|---|---|
| | | Time spent | Amounts | |
| System construction | | | | |
| Informing candidates | | | | |
| Preparing files | | | | |
| Registering candidates | | | | |
| Organizing groups | | | | |
| Designing contents | | | | |
| Loading courses | | | | |
| Supervising learners | | | | |
| Organizing exams | | | | |
| Evaluating learners | | | | |
| Total | | 100% | | |

Table 1: Example of a grid for online learning tasks and their rating modes

The following grid, developed in the same framework, offers a more detailed list of activities and their attributions to different actors of an online learning system.

---

[9] Arnaud Coulon et Michel Ravailhe. « Les couts de la formation ouverte et à distance : première analyse » [http://sup.ups-tlse.fr/documentation/docs/fich_118.pdf]

| N° | ACTIVITY TITLE | A | B | C | D | E | F | G | H | I | J | K | |
|---|---|---|---|---|---|---|---|---|---|---|---|---|---|
| | **ELEARNING CONTRIBUTORS** | | | | | | | | | | | | |
| A01 | Studying the project feasibility | | | | | | | | | | | | A: Pedagogical responsible |
| A02 | Setting prototypes and testing them | | | | | | | | | | | | |
| A03 | Developing and organizing infrastructure | | | | | | | | | | | | B: Project leader |
| A04 | Endorsing the change | | | | | | | | | | | | |
| A05 | Promoting eLearning | | | | | | | | | | | | C: Coordinator/animator |
| A06 | Managing eLearning input and output | | | | | | | | | | | | |
| A07 | Instructing the files | | | | | | | | | | | | D: eTutor |
| A08 | Creating and validating contents | | | | | | | | | | | | |
| A09 | Adapting, integrating and loading online | | | | | | | | | | | | E: Teacher |
| A10 | Acquiring contents | | | | | | | | | | | | |
| A11 | Training actors | | | | | | | | | | | | F: ADM |
| A12 | Information watching | | | | | | | | | | | | |
| A13 | Ensuring reporting | | | | | | | | | | | | G: Administrative relations |
| A14 | Regulating the eLearning system | | | | | | | | | | | | |
| A15 | Animating the eLearning system | | | | | | | | | | | | H: Administrative staff (example: accountability) |
| A16 | Publishing and updating documents on server | | | | | | | | | | | | |
| A17 | Server administration | | | | | | | | | | | | I: VAP committee () |
| A18 | Improving, adapt existing resources | | | | | | | | | | | | |
| A19 | Receiving/welcoming learners | | | | | | | | | | | | J: CNED |
| A20 | Planning eLearning scenarios | | | | | | | | | | | | |
| A21 | Accompanying learners | | | | | | | | | | | | K: Learners |
| A22 | Preparing and animating groupings | | | | | | | | | | | | |
| A23 | Feeding the platform | | | | | | | | | | | | |
| A24 | Correcting and publishing exercises | | | | | | | | | | | | |
| A25 | Preparing tests | | | | | | | | | | | | |
| A26 | Correcting, validating and publishing results | | | | | | | | | | | | |
| A27 | Accompanying learners' graduation | | | | | | | | | | | | |
| A28 | Assessing and reporting on the experience | | | | | | | | | | | | |

ELEARNING BASED ACTIVITIES

Table 2: eLearning Activities and their assignments to actors (Source: Coulon & Ravailhe, 2003)

In another example, published by Druhmann & Hohenberg[10], different process and instruments of eLearning economic assessment are proposed. Firstly, a cost structure was specially tailored to the local blended learning program. "This was necessary, as so far no standard cost structure has been developed on the basis of the various approaches in the literature. Cost-generating activities were initially identified on the basis of the teaching-learning processes, and the associated cost types defined – structured according to the phases of design, use and development of the local blended learning measures"[11]. Three types of costs are identified:

1. Development costs: they cover one-time costs for initial development of the system;
2. Operating costs: they cover annual costs around training activities;
3. Cost of services: they cover maintenance and further developments per year for improving learning quality.

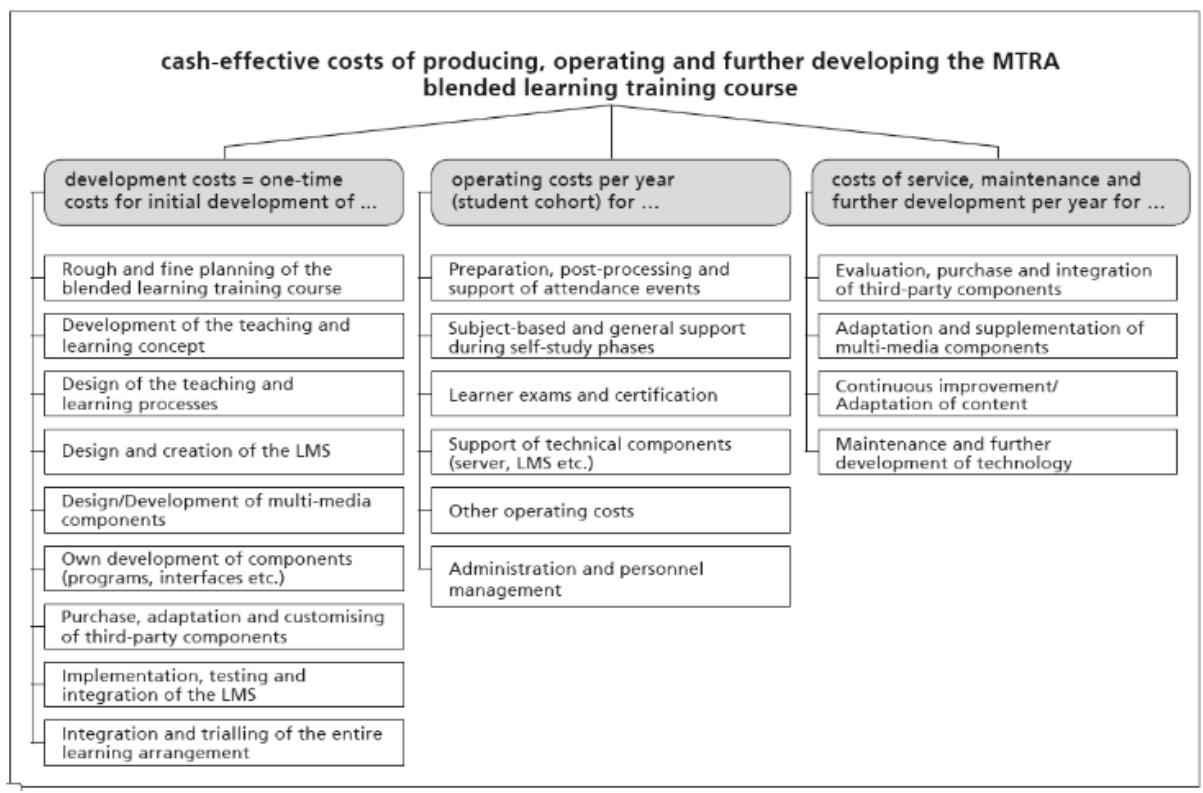

Table 3: Pragmatic cost model for determining the payment-related expenses for production, operation and development of a local training course (Source: Druhmann & Hohenberg, 2014).

Generally, to estimate the actual costs of eLearning activities, almost all methods follows an analysis approach distinguishing costs, or charges, along two different principles[12]:

- Direct charges and indirect charges;

---

[10] Druhmann, C., & Hohenberg, G. (2014). Proof of economic viability of blended learning business model. In Facebook Mediated Interaction and Learning in Distance Learning at Makerere University (ERIC Clearinghouse, 2014, p. 70-78).
[11] Op. Cit.
[12] For an application case study of the ABC method, see: Arnaud Coulon and Michel Ravailhe. "The costs of open and distance learning: first analysis" [http://sup.ups-tlse.fr/documentation/docs/fich_118.pdf]

- Fixed charges and variable charges.

|  | Fixed charges<br>(or structural charges)<br>They are constant even when the activity varies but can produce threshold effects: for example, a training room will be adapted to a determined number of trainees | Variable expenses<br>(or loads ' activity)<br>They vary with the level of activity: based on volume training schedule, the number of trainees, the number of groups |
|---|---|---|
| **Direct charges**<br>They are directly related to the training action | • Staff costs: Project management, engineering of educational resources<br>• Purchase / production of digital educational resources (licenses / maintenance)<br>• Purchase tools - learning platform (LMS) - or rent (if the cost is fixed e regardless of the number of ' learners), Maintenance<br>• Local: Training room, computer room / media resource center 's dedicated to training<br>• Communication costs | • Staff costs: trainer, accompanist, tutor*, educational coordinator: course prescription, disciplinary tutoring* and learning methodology support (synchronous* and asynchronous* time) and the course, animation of pedagogical sequences in presence or in tele-presence - type of class Virtual<br>• Rental of unit licenses for multimedia educational resources<br>• Educational resources and documentation or materials given to trainees (duplication, dissemination) |
| **Indirect charges**<br>They are common to several activities of the training organization<br>They are carried out according to distribution keys.<br>The training action bears "its fair share" of the overall structural cost | • Structural costs: Support functions (administrative and technical services), supervision (director and educational manager); Cost of non-educational premises<br>• Depreciation of equipment (eg: IT)<br>• Staff training |  |

Table 4: Types of charges for the provider of an online learning[13]

From this non-exhaustive grid, the pilot team of the blended learning project should set the cost drivers (indicators) in order to be able to measure the performance of an activity and represent its consumption of resources. The cost driver can be compared to a work unit (according to the general accounting plan of the institution) which is the unit of measurement used in particular to calculate the cost of providing a service. The following grid gives some examples:

---

[13] Financement et mise en œuvre de la FOAD : Vade-mecum des bonnes pratiques. www.una-univ-bordeaux.fr/Download/News/Info/document/233.pdf

| Activities / service | Examples of cost drivers |
|---|---|
| Registering learners | According to the number of candidates |
| Creating learning content | According to the number of pages / recording times (audio-video), etc. |
| Tutoring | According to the number of students / period / duration / status |
| Evaluating homework | Depending on the number of assignments / types of support / frequency, etc. |
| … | … |

Table 5: Examples of drivers to measure the cost of activities of online learning

# 3    PEDAGOGICAL MODELS OF BLENDED LEARNING

It is sometimes a difficult task to find the appropriate method to start a blended education. However, teachers / trainers today have a myriad of options and a variety of solutions to introduce the use of ICT in their lessons. The main thing now is to know "how to get started or develop the mixed or blended approach to training" rather than wondering whether "we should get started or not in distance learning using digital technologies".

It is important to remember also that between full face-to-face teaching and entirely distance learning, there are several levels of integration of ICT in educational practice (ICTE). We often measure the level of blending education in terms of the proportions between residential education (full face to face or on site) and full virtual distance education. There are many combination alternatives, which explains the complexity of choosing a specific formula to suit a particular training intended for a given category of learners.

Finding the right "combination" or compromise between the two types of activity is part of a typology of blended learning that Serge Leblanc (University of Montpellier) has developed on the basis of face-to-face support by digital distance processes:

- **Face-to-face enriched by multimedia supports**: use in class by the trainer and / or the learners of presentation tools or multimedia resources. For example: Use of slide shows, exploitation of resources from the Internet (manufacturer's documentation, installation videos).
- **Face-to-face improved by upstream or intersessional work**: All teachers and students have an electronic mailbox. Before and after the lesson, the teacher provides students with a number of resources that they can reach remotely.
- **Alternate face-to-face**: Setting up courses which alternate face-to-face and distance learning.
- **Lightened or reduced face-to-face**: Most of the training is in face-to-face. Some distance work is replaced by self-study, individual or group work.

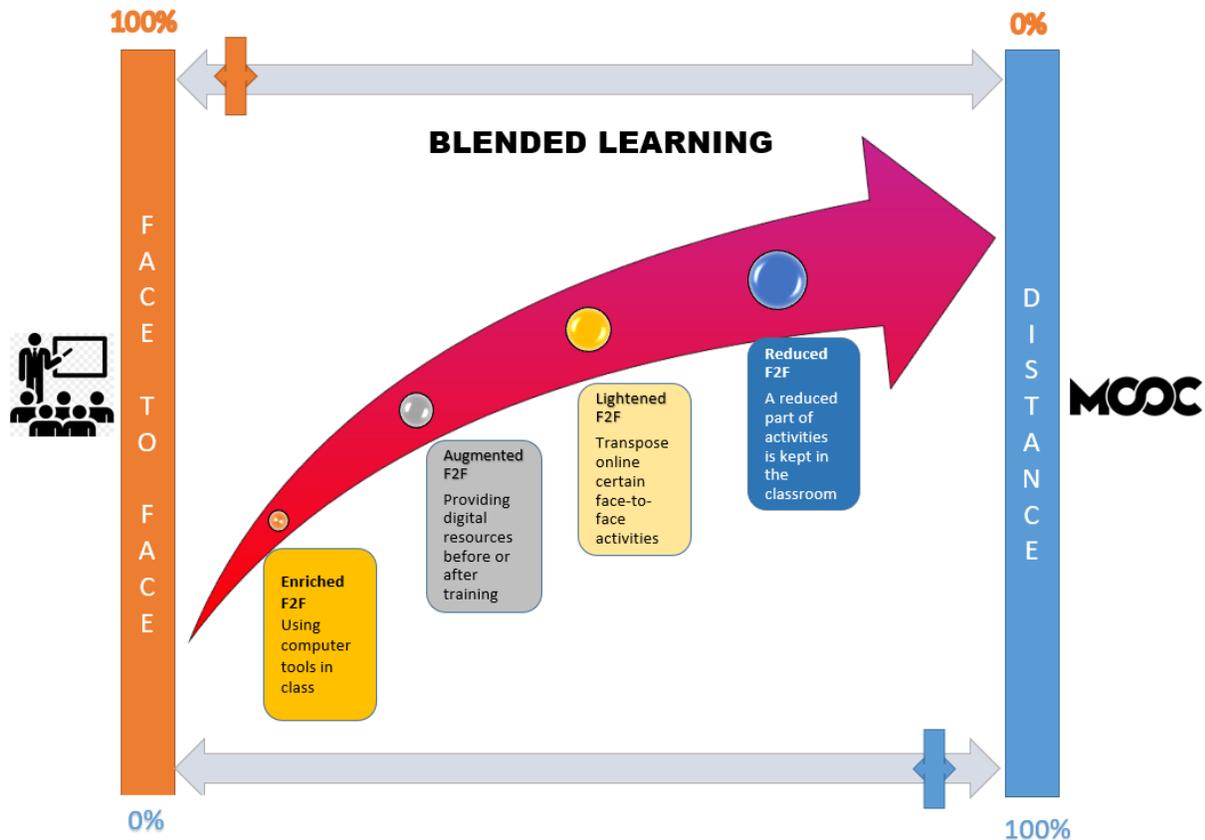

Figure 1: Blending levels of learning, from full face-to-face to full online

Daniel Peraya[14], for his part, established six types of configurations around the activities of teaching and blended learning:

- **Type 1**: Content-oriented "teaching" configuration, characterized by support for face-to-face lessons and the provision of essentially textual resources ("the stage").

- **Type 2**: Content-oriented "teaching" configuration, characterized by support for face-to-face lessons and the provision of numerous multimedia resources ("the screen").

- **Type 3**: Configuration "teaching" oriented course organization by the use of management tools and sometimes tending towards the integration of relational and reflexive objectives ("the cockpit").

- **Type 4**: "Learning" configuration centered on supporting the knowledge-building process and on interpersonal interactions ("the crew").

- **Type 5**: "Learning" configuration centered on the opening of the learning system to external resources during the course and promoting the freedom of choice of the learners in their learning path ("public space").

- **Type 6**: "Learning" configuration characterized by the exploitation of a large number of technological and educational possibilities offered by blended systems ("the ecosystem").

---

[14] PERAYA Daniel et. al. « Typologie des dispositifs de formation hybrides : configurations et métaphores ». AIPU. Quelle université pour demain ? Mai 2012, Canada. pp.147-155, 2012

The advent of ICTs in the educational context and the transformations they have brought about in teaching methods and training offers have not been without constraints. These constraints have strongly affected the governance models of universities and training institutions to the point of pushing them towards new strategies of alliances, consortia and autonomy. The two constraints that have most marked bimodal and blended distance education are undoubtedly the legal framework and the business model which could be very different from conventional educational models. The promotion, in terms of costs, of distance learning practices and the recognition of the resulting diplomas, were among the stumbling blocks faced by those involved in educational innovation and online learning.

# 4 A BLENDED LEARNING PROJECT: TEAMS AND GUIDELINES

Blended solutions, as well as full virtual, are now within everyone's reach, provided you prepare for it and have the prerequisites to succeed.

The implementation of blended learning can, in fact, be assimilated to project management which requires work teams and documents for coordination, monitoring and assessment.

## 4.1 THE WORK TEAMS OF A BLENDED LEARNING PROJECT

A blended learning project is a team-project involving teachers and students as well as technicians and administrators responsible for financial and human resources matters. Each intervenes according to his role and his professional profile as defined in the preliminary project file established by the institution.

Without this being a general rule, a blended learning project can be supported by three types of teams: a project team (PT), a steering committee (SC)* and a monitoring committee (MC). This organization is given as an indication. Each institution can adapt to its own situation.

### 4.1.1 Project team

The project team (PT) is made up of resource persons designated according to their profiles, roles and commitments in pedagogical innovation by the online learning. The project team studies the needs of the university and prepares an application file to respond to any type of call or setting up blended learning projects. It is strongly recommended that this team be versatile, made up of people with educational, technical, financial, legal and human resource management skills. Its main task is to prepare the most complete project file by providing as much information as possible about its feasibility.

### 4.1.2 Steering committee

A steering committee (ST) * is formed as soon as a blended learning project begins to be discussed within the institution. It is mainly composed of the extended project team of one or more experts representing one or more key partners (i.e. AUF in the case of projects proposed by university members of CONFRASIE). This joint committee oversees the smooth running of the project according to the objectives and established operating procedures.

The steering committee* writes regular reports on the conditions of the activities during the support period. Its powers, set out in a framework agreement between the parties concerned, are defined by mutual agreement between the partners.

### 4.1.3 Monitoring committee

A monitoring committee is made up of the project team and the steering committee* to which are added potential partners from the host university who are associated with the blended learning project (other universities, private sponsors, structures research, industry, etc.).

The main task of the monitoring committee is to watch over the good progress of the project towards its educational objectives. He intervenes when necessary on questions of a strategic order or relating to the governance model of the online learning system in place.

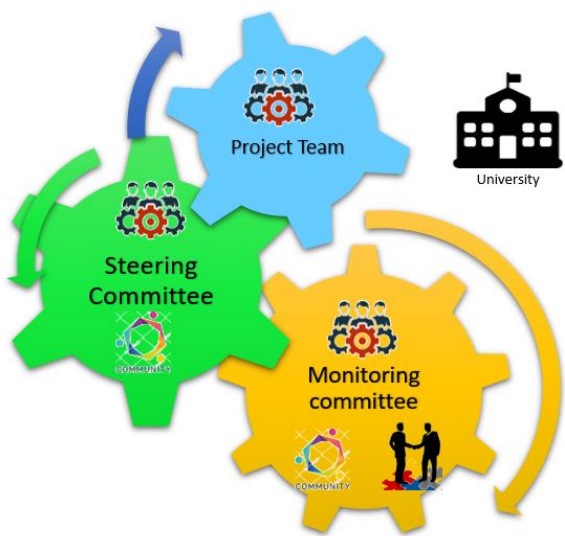

Figure 2: Committees around a blended learning project

## 4.2 THE TECHNICAL DOCUMENTATION OF A BLENDED LEARNING PROJECT

A blended learning project generally follows an operating logic defined in several types of working documents:

- A pre-project proposal form,
- A pre-project description form,
- A project charter document,
- A feasibility study report,
- A project plan or roadmap,
- A quality assurance report,
- A project assessment report,
- A project closure & delivery report, etc.

This process is not static. It adapts to the reality of each institution and to the complexity nature of the project being implemented.

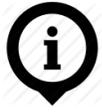 It is not mandatory to include all the documents indicated in a blended learning project file. The process of a project application file can possibly vary according to several criteria related to the reality of the context and the state of experience of each institution.

This Guide offers a series of documents that can respond to several scenarios for deploying blended learning. These documents are classified into three categories providing for a project to be carried out in three steps:

### Step 01: PRE-PROJECT:

1. A pre-project presentation form: institutional framework (Appendix 01);
2. A pre-project presentation form: means, objectives & requirements (Appendix 02);
3. A standard format of a project opportunity memorandum (Appendix 03);
4. A project plan or roadmap (Appendix 04);
5. A technical framework for a blended learning project (Appendix 05);

### Step 02: IMPLEMENTATION:

6. Monitoring survey form: project manager (Appendix 06);
7. Monitoring survey form: content designer* (Appendix 07);

### Step 03: ASSESSMENT & DELIVERY:

8. Pre-defined learning assessment grid: learner (Appendix 08);
9. Customized learning assessment survey: learner (Appendix 09);
10. Customized learning assessment survey: tutor* (Appendix 10);
11. Record form and project closing report (Appendix 11).

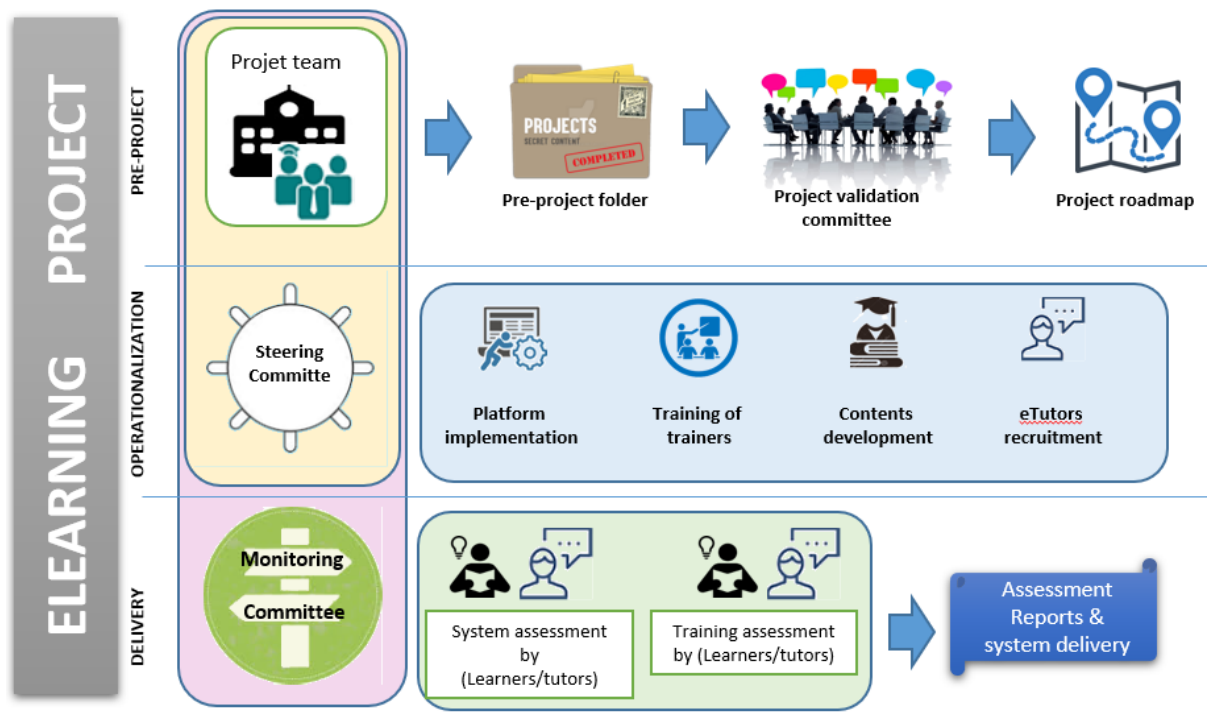

Figure 3: Master plan of a blended learning project

Let us also remember that an online learning system is defined by Marc Weisser "as an articulation of heterogeneous, material and symbolic elements (Charlier & Peter, 1999; Weisser, 2007), and as a set of means implemented with an explicit aim, at least in the mind of its designer (Meunier, 1999). It is through it that the teacher tries to foresee and mark out the learning path* that he offers to his learners, under the influence of his didactic or pedagogical choices"[15].

## 5    WHY & FOR WHOM THIS GUIDE?

This guide is proposed as a theoretical, technical and methodological reference document for open and distance learning (online learning). As such, it provides benchmarks for academic institutions in their blended learning projects. It can serve both promoters of blended learning projects as well as teachers and trainers wishing to introduce a new pedagogical modality of teaching using ICT.

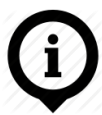 It is important to recall that this guide addresses the blended aspect of transmitting knowledge, regardless of the professional or academic contexts. The taxonomy used therefore alternates between "learning" and "training" depending on the situation. In general, "training" concerns the professional career of a company staff, while "learning" and "teaching" concern the educational system in educational institutions.

This guide is also intended to serve as support for e-training managers but also as methodological help for the teams of trainers involved in this type of approach. This Guide can therefore be useful from a strategic point of view for those responsible in charge of innovating the educational policies of their institutions. It can also help - at an operational and practical levels - in the design and implementation of online learning systems for the blending of existing face-to-face trainings.

---

[15] Marc Weisser, « Dispositif didactique ? Dispositif pédagogique ? Situations d'apprentissage ! », Questions Vives [En ligne], Vol.4 n°13 | 2010, mis en ligne le 26 janvier 2011, consulté le 12 mars 2017. URL : http://questionsvives.revues.org/271

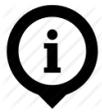 It is also important to remember that this Guide only covers the part of blended learning delivered online by technological means. The face-to-face part remains the responsibility of the teaching team of the partner university. The face-to-face mode in a blended learning project is limited in this guide to the training of trainers' workshops that would be given to the project team.

It is obvious that this Guide is not intended to be exhaustive and does not seek to be consistent with all situations which are very diverse in nature. It also does not seek to go into the details of the technical aspects, educational procedures, the creation of specifications, etc. Many resources, tools and services available on the Internet can later supplement this Guide which is ultimately limited to the principles and main lines of open and distance learning, especially in the choices for free software, free educational resources and Creative Commons licenses.

# 6   THE GUIDE STRUCTURE

Taking into account the criteria and conditions supporting the process of accompanying an online learning system, the structure of this Guide follows a modular organization arranged in three parts:

- **Part 01**: preliminary analysis of the context of the blended learning system to be implemented. The Guide lays down the criteria and conditions under which blended learning would be carried out (human, technical, educational, financial, etc.)
- **Part 02**: development of skills linked to blended learning. The Guide offers all the technical and pedagogical skills on which designated trainers should receive a prior training to master all of the activities and pedagogical objectives of blended learning. This concerns skills relating to the manipulation of online learning tools (including platforms and related applications), the design of educational content (structuring, scenarios), tutoring* (support), assessment (impact), etc. The skills and courses offered are chosen from multiple standards, in particular that of AUF. These essential skills would enable the implementation of the online learning system by designing and integrating educational resources and learning activities on the new online learning system;
- **Part 03**: implementation, experimentation, assessment and validation of the online learning system. To this end, the Guide proposes the conditions under which the experimentation, assessment, validation and delivery of the online learning system would take place.

The Guide will conclude with recommendations for the pooling of blended learning within university institutions and its potential evolution towards fully distance learning (i.e. Mooc).

The blended learning project would be broken down into 03 essential phases:

- **Phase 01: Design** and implementation of the online learning system;
- **Phase 02: Operationalization** of the online learning system;

- **Phase 03: Experimentation** and validation of the online learning system.

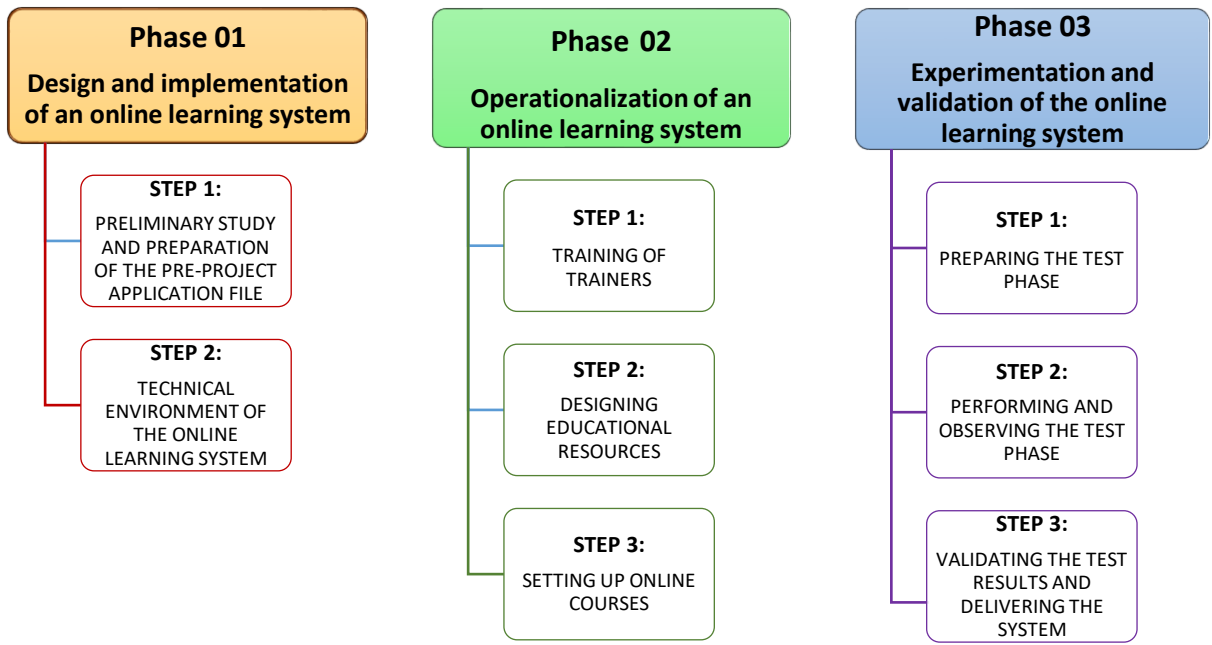

Figure 4: The three phases constituting the main sections of the Guide

**REMINDER**: The organization of the guide follows a procedure independent of any institutional context. The phases and steps that constitute its parts are proposed as modalities that each institution or project leader can adapt to their own contexts and peculiarities. The modularity of the guide thus allows the phases and steps to be arranged in an order that would best suit a specific situation.

# PHASE 01 - DESIGN AND IMPLEMENTATION OF AN ONLINE LEARNING SYSTEM

Any blended learning project must rely on "robust" design criteria to keep safe against a number of start-up pitfalls. These criteria are identified through a preliminary analysis of existing resources.

To commit to a blended learning project, the institution should first provide a pre-project proposal which establishes an inventory of the means and resources available (**Appendix 01**). This proposal can take the layout of a descriptive form which explains the relevance of the project to a validation team (a sponsoring partner) who could retain, reject or call for the revision of the proposal. This step is optional and depends on the organizational strategy of each institution which can go directly to the subsequent steps.

Recall that the assessment of a blended learning project is done according to criteria commonly observed in project management, defined on the basis of the principles of relevance, consistency and feasibility of the project (see next information box):

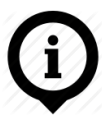

**Relevance of the project**: these are indicators which take into account the scientific, socio-economic and institutional environment (governmental and inter-governmental) in which the training takes place. Three elements can, in particular, be taken into consideration: priority objectives of university programming; potential for recruiting learners; potential for professional integration and postdoctoral studies within the framework of a university career.

**Coherence of the project**: the assessment of the coherence of the project should lead to highlighting the concordance, at the level of the organization and structuring of the project and its constituent elements, between the previously defined objectives, the means to be implemented work and expected results. Three elements can, in particular, be taken into consideration: training program with regard to the knowledge and skills targeted at the end of the training; steering or monitoring committee\*; communication plan and target audiences.

**Feasibility of the project**: the feasibility assessment must take into account, with regard to the means implemented and foreseeable events, the progress, difficulties and obstacles that may affect the achievement of the training and the objectives initially defined. Three elements can, in particular, be taken into consideration: budget estimate and search for financial empowerment; teacher training plan of the institution carrying out the training project; partnerships.

The purpose of this first pre-project file (project proposal sheet and inventory form) is to determine the criteria and conditions under which blended learning can be successfully carried out. It is also a question of identifying the indicators likely to determine which type of blended learning can be in conformity with a particular training module which is part of a specific training offer in a determined university context.

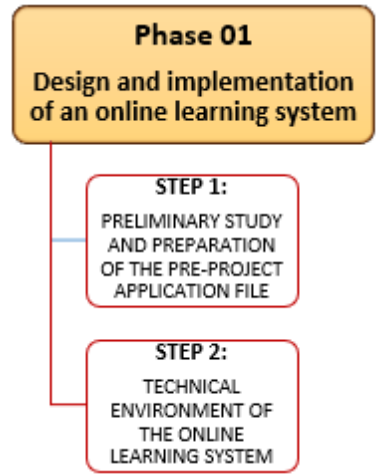

This phase of assembling a project file for an online learning takes place in two stages:

1. **Step 1**: Preliminary study and preparation of the application file for the pre-project application file;
2. **Step 2**: Technical environment of the online learning system.

# 7  STEP 1: PRELIMINARY STUDY AND PREPARATION OF THE PRE-PROJECT APPLICATION FILE

A new project usually starts with a preliminary study of existing means and resources and the identification of needs and requirements following which goals are set, the actors are identified and means are unlocked.

Any pre-project is also subject to a prior assessment by an experts committee to study its consistency, relevance and compliance with the current economic situation. Once validated, the pre-project should immediately be consolidated by a report that takes form of a "realization plan" also called "project roadmap" associated to an operational plan (diagram).

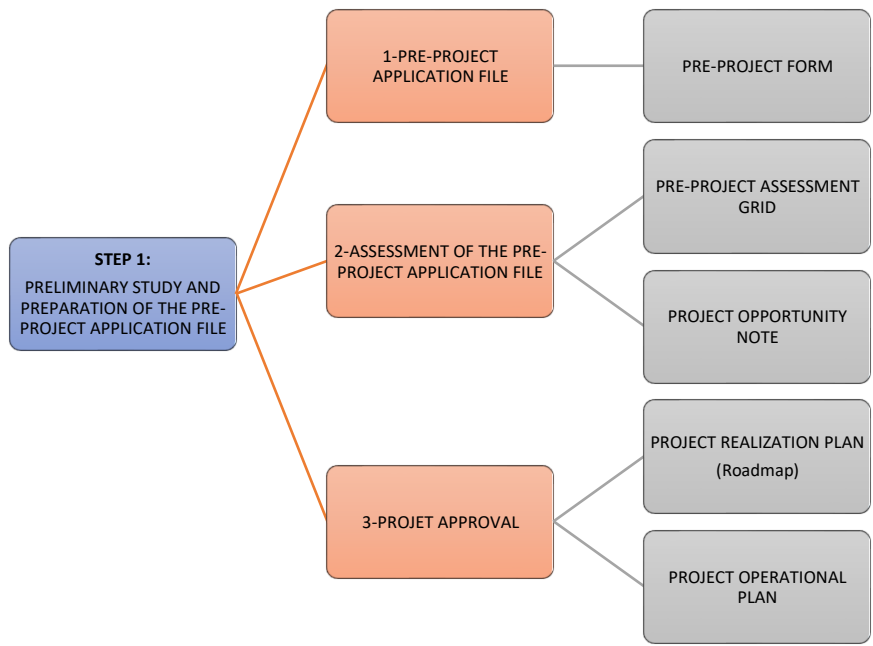

Figure 5: General breakdown of a blended learning project

## 7.1 PREPARING THE PRE-PROJECT APPLICATION FILE

The proposal for supporting a blended learning system must begin by presenting an inventory of the environment where the online learning portion of the project will take place (**Appendix 01**).

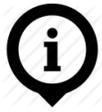

> This step varies according to the contexts and the objectives of the project. Appendix 01 provides a sample form to be used in the case of a project to be evaluated by a validation experts committee. The initiators of blended learning projects can do without it (or be inspired of) in setting up a blended learning project in their institutions, as a standalone initiative or with national or international partners.

A pre-project form is designed to provide information that outlines the global framework of a blended learning project.

The purpose of this form is to convince all involved parties of the general interest of the project and its feasibility. The form writer must therefore act as a "seller" of an idea and highlights the benefits that his institution can derive from the project. He must appropriately declare the needs and objectives of his institution by implementing blended learning. He must also justify the existence in his university of a minimum potentials (human, technological, financial, etc.) that can help the flexible and rapid implementation of a blended learning.

Among the information that the pre-project form must provide:

- Presentation of the project objectives;
- Information on the project leader and his partnership framework;
- The educational characteristics of the project;
- The technical characteristics of the project;
- The project governance model;
- The language policy of the project;

## 7.2 ASSESSMENT OF THE PRE-PROJECT APPLICATION FILE

The assessment of the application file generally consists of checking the relevance of the following facts:

- If the project team has initial skills related to the online learning;
- If the project team has prior educational resources (content, activities, assessments, etc.) consistent with online learning usage;
- If the host institution for the blended learning has a high-quality technological infrastructure (IT and telecommunications) necessary to manage the online learning part of the blended learning (student equipment, public Wi-Fi access, VPN, ADSL network);
- If the institution has clearly defined the educational needs to be met in order to administer blended learning from the point of view of administrative and technical stewardship, educational coordination and budgetary management;
- If the institution has well defined an economic model specific to the online learning part, envisioned in accordance with the institution rules and the financial legislation in force in the country;
- If the project team has clearly defined the objectives to be achieved, planned the actions to be carried out and identified the resources to be mobilized;

- If the project team has specified the roles and attributions of each of the members involved in supporting the project (project manager, legal advisor, educational manager, technical manager, designers of educational resources, tutors*, etc.).

The pre-project validation committee uses this form to establish a project validation or rejection decision. It can possibly communicate its decision by producing two kind of documents:

1. A pre-project assessment grid;
2. A project memorandum of opportunity.

### 7.2.1 Pre-project assessment form [Appendix 02]

The validation committee can draw up its report using an assessment matrix in which it signals the presence of criteria in favor of the project. These criteria are clustered under the following categories:

- General environment of the project;
- Declared needs of the project;
- Declared objectives of the project;
- Prior skills in favor of the project;
- Existing educational resources recoverable in the project;
- Technological infrastructure available favorable to the project;
- Stakeholders identified in the course of the project;
- Financial resources mobilized for the project (economic model).

Acceptance of the proposal is confirmed if a predetermined percentage of positive criteria is reached. This rate is proposed as a minimum prerequisite for an institution to be supported in a blended learning project.

This rate can nevertheless be readjusted (up or down) in the following two cases:

- It can be revised upwards if the online learning (the distance side of the project) is set to more or equal to 50% of the entire module or training program proposed to be conducted in blended mode. A high percentage of online learning requires a good predisposition and "robust" prerequisites on the part of the host institution to ensure the monitoring and control of the project;
- It can be revised downwards if the distance portion of the project is less than 25% of the entire module or training program proposed to be conducted in blended mode. A small percentage of online learning requires less predisposition and less prerequisites on the part of a host institution to ensure the monitoring and control of the project.

A matrix model is proposed as an example in (**Appendix 02**). It is a working tool for collecting the quality indicators of a project.

### 7.2.2 Pre-project opportunity memo [Appendix 03]

The opportunity note is a decision-support tool. It provides an initial assessment of the relevance of the project and makes it possible to confirm or deny the advisability of the commitment in supporting the blended learning pre-project.

In view of this note, it will be decided to stop the project, to continue it or to call for its revision. In the latter case, the assessor must give specific recommendations to optimize the conditions for acceptance of the project in an enriched version which could be re-submitted later.

The project opportunity note should emphasize the following points:

- The training needs, even its priority within the institution, as defined by the project team;
- The situation of the host institution and the possible limits for a blended learning project;
- The benefits of blended learning for the concerned institution;
- Key success and potential risk factors in the institution to complete such a project;
- First estimate of the costs of such a project (economic model);
- Decision: make the choice to launch the project or not.

For medium and large projects (several partners and large numbers of learners), the opportunity note is normally essential before embarking on a more in-depth feasibility study.

## 7.3 THE PROJECT APPROVAL

Once approved, the project team should initiate an initial working meeting to finalize the project documents. Two documents can be produced at this stage in order to confirm the execution of the project and fix it in space and time:

1. A project realization plan (a roadmap);
2. An operational plan (Gantt chart).

### 7.3.1 Project realization plan (Roadmap) [Appendix 04]

It is essential, after the validation process, that a detailed action plan* be carried out. It is sort of a project roadmap that needs to be developed, documented, and agreed to by stakeholders.

The roadmap answers the questions "when", "who" and "how" are we going to achieve the expected results. It describes the organization of the work, identifies the specific actors and activities assigned to them and sets the timetable and the resources necessary to succeed in the planned action strategy. Ultimately, the roadmap defines how the project will be executed, monitored and controlled, and then closed.

A roadmap for a blended learning project must cover as a priority the economic model of the project, the stakeholders, their roles and responsibilities, the educational resources to be digitized and exploited, the support, monitoring and reporting mechanisms, an implementation schedule, etc.

The general structure of a roadmap can be built around the following points:

- Type of blended learning model to implement;
- Methodology: how to lead the network of digital experts;
- Stakeholders, roles and responsibilities;
- Digital environment of blended learning;
- Implementation responsibilities;
- Training of supervisory staff;
- Agenda management;
- Risk management;
- Cost management (business model);
- Monitoring and reporting;
- Closing reports and deliverable.

A detailed model of an online learning project roadmap is given in **Appendix 04**.

### 7.3.2 Project operational plan

An operational plan is a document by which the person in charge of a project sets the objectives he wishes to achieve within specified deadlines. The operational plan must be consistent with the action plan*. The two are to a certain extent essential guides which offer the project leader an overall view of the project and the stages of its implementation.

The operational plan is often produced in a graphical form (i.e. Gantt chart) which allows a project manager to monitor on a mass plan (global matrix) the progress of all the stages or milestones that make up the project.

Using a Gantt chart allows you to set up and manage complete projects according to a standardized and proven methodology. With a Gantt chart, it is easy to add or delete tasks, define or modify their duration, chain tasks. Special computer applications could also be used for this task.

## 8 STEP 2: TECHNICAL ENVIRONMENT OF THE ONLINE LEARNING SYSTEM

The characteristics of the project, detailed in the roadmap (realization plan), provide the elements of the hardware configuration of the technical environment of the blended learning system.

As a reminder, the intervention of the assessment committee is carried out only on aspects related to the distance part of the blended learning project. The organization of the face-to-face part is under supervision of the institution. Within the framework of this Guide, only the face-to-face trainings on the online learning techniques, delivered to the project team, will be considered.

In this step 2, three essential choices are to be determined:

1. The choice of an educational platform;
2. The choice of a hosting solution;
3. The choices in implementing and configuring the platform.

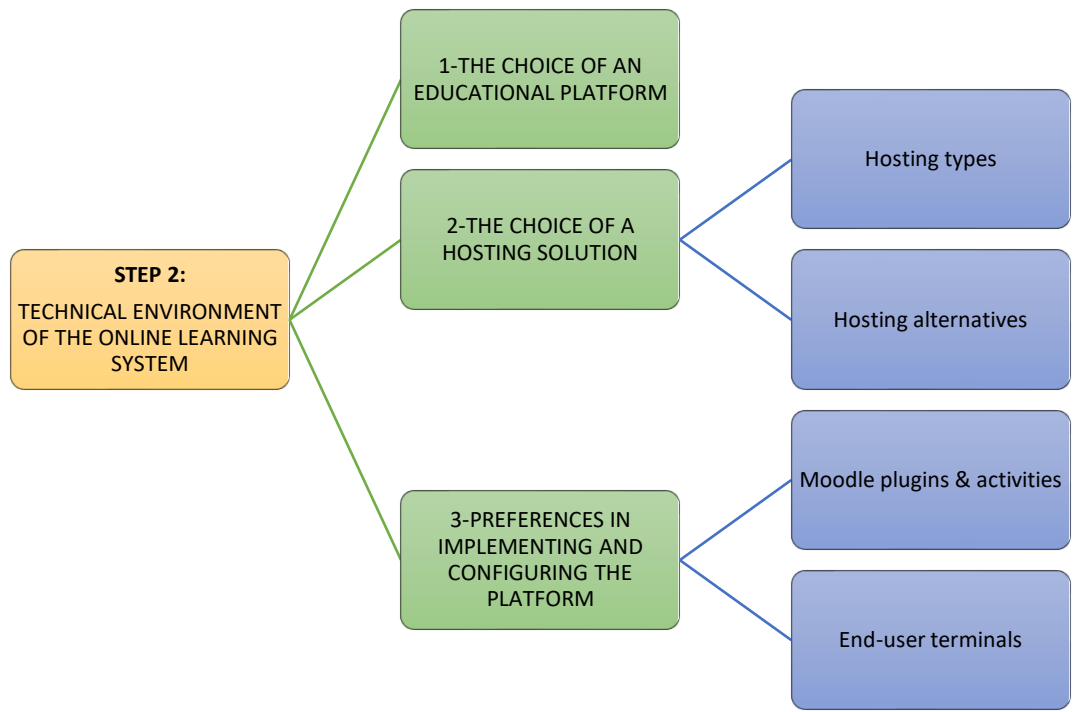

Figure 6: Elements of choosing a technical environment for blended learning

## 8.1 CHOOSING AN EDUCATIONAL PLATFORM

The choice of an educational platform is crucial, because it provides the blended learning project with an innovative pedagogical engineering model which replaces a current practice rather assimilated to an inactive online course distribution.

The platform choice is often made between a CMS (Content / Course Management System), an LMS (Learning Management System) or an LCMS (Learning Content Management System). The difference between LMS and CMS lies in the fact that the LMS has an educational implication, and therefore, in addition to communication and content delivery tools, allows services to monitor not only student activity, but also assessment or collaborative work.

The market for educational platforms is very prolific in solutions as important as each other, but the trend has been overwhelmingly oriented towards a tool in the free software category, in this case the Moodle platform[16]. Moodle benefits of a large international community open forums[17] and online assisting resources[18].

---

[16] Moodle downloads: https://download.moodle.org/
[17] Moodle forums standards: https://moodle.org/mod/forum/index.php?id=5
[18] Moodle documents: https://docs.moodle.org/39/en/Resources

According to usage statistics, updated online in April 2020[19], Moodle is installed more than 154,000 times in more than 240 countries, with more than 24 million courses and more than 200 million users.

In addition to being a free software, Moodle reproduces the conditions for an action-based learning compliant with the new pedagogies aligned with constructivism* and socio-constructivism* theories. Its activities and complementary modules (plugins) are conducive to what is now known as active and dynamic group teaching. It is therefore highly recommended as a choice of an educational platform for any online learning project. The guide proposes proceeding with the techno-pedagogical specificities of Moodle Learning Management System.

The choices that remain to be made concern first the implementation of the most recent version of Moodle to take advantage of its latest features, then the choice of a hosting solution to be determined between shared, private or virtual hosting.

## 8.2 Choosing a hosting solution

The choice of hosting Moodle platform should not be overlooked. It should be considered from a technical point of view as well as from managerial and administrative perspectives.

On the technical level, it is important to know the characteristics of the different types of hosting offers that are made today on the Internet according to very differentiated offers ranging from simple shared hosting to dedicated hosting, to personalized virtual servers (VPS), to servers in the clouds (Cloud Computing).

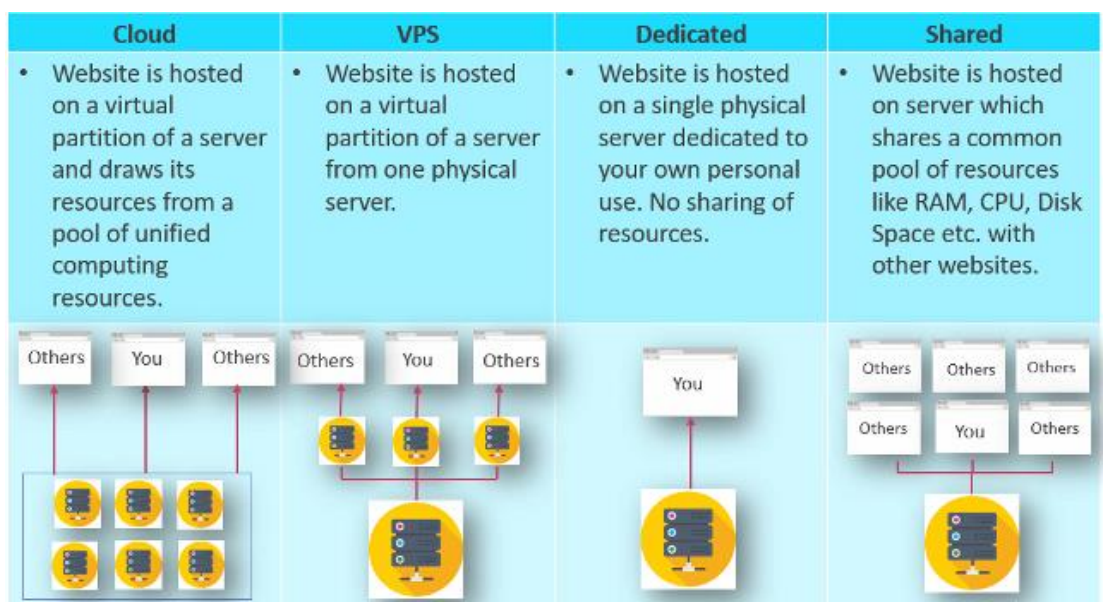

Figure 7: Different web server hosting models (Source: https://www.webhostingsecretrevealed.net, 2019)

---

[19] Moodle statistics. https://stats.moodle.org/

It should be remembered that Moodle is developed in a LAMP environment: Linux as operating system, with Apache as server, MySQL as database system and PHP as programming language. For hosting Moodle, the platform works also with Unix, FreeBSD, Windows and also MAC OS X and Netware and all other systems that have a PHP web server (version 4.1.0 or more) and a basic management system of MySQL or PostgreSQL data.

On the managerial and administrative levels, it is essential to make a choice that would best suit the logistical and organizational capacities of the institution to manage the online learning by itself or to call on external actors.

### 8.2.1 Types of Internet hosting

▷ **Shared hosting: the cost-effective solution**

Shared hosting involves using the same server to host multiple websites. This allows the host to reduce its maintenance costs and offer a low-cost solution. It is the most inexpensive and widespread solution in the world, since some servers can host several hundred websites at the same time generally reserved for personal websites. Shared hosting is a suitable solution for sites with low or medium traffic (less than 1000 visits per day).

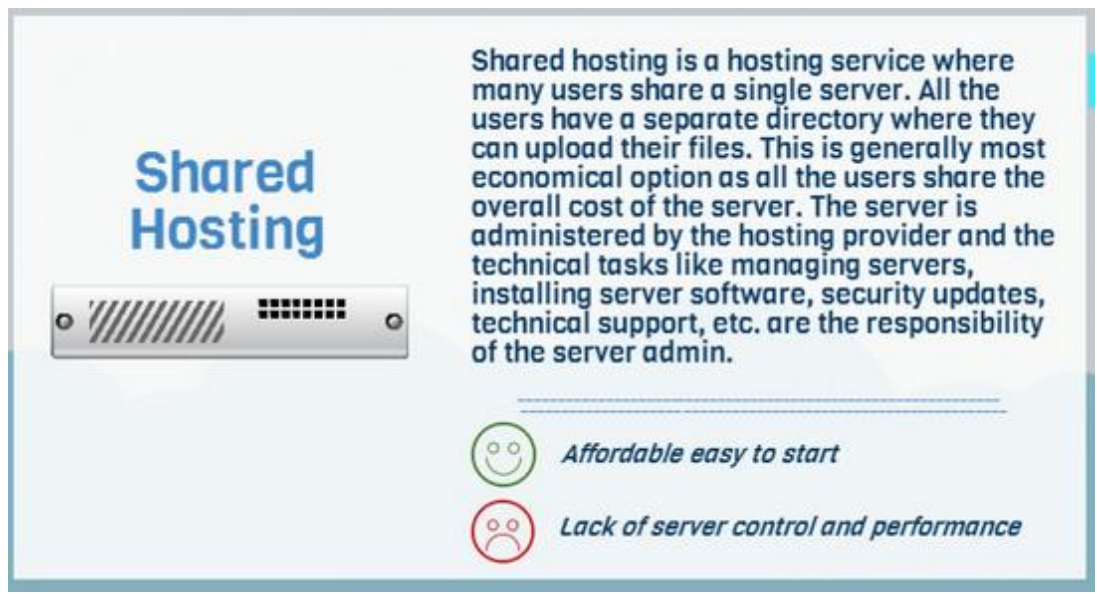

Figure 8: Shared hosting (Source: Visually & Rockcontent)

| Advantages | Disadvantages |
|---|---|
| • Least expensive hosting<br>• Maintenance, security and backup are in principle provided by the host | • Free or entry-level solutions, generally reserved for personal websites<br>• Solution not suitable for websites with high traffic or requiring a specific configuration. |

▷ **The dedicated server hosting: the controlled solution**

Hosting on a dedicated server allows you to have your own machine entirely and exclusively dedicated to a website. This configuration is most used for hosting high traffic sites. Server configuration is the sole responsibility of the site owner, who often calls on a professional to ensure optimal performance and security.

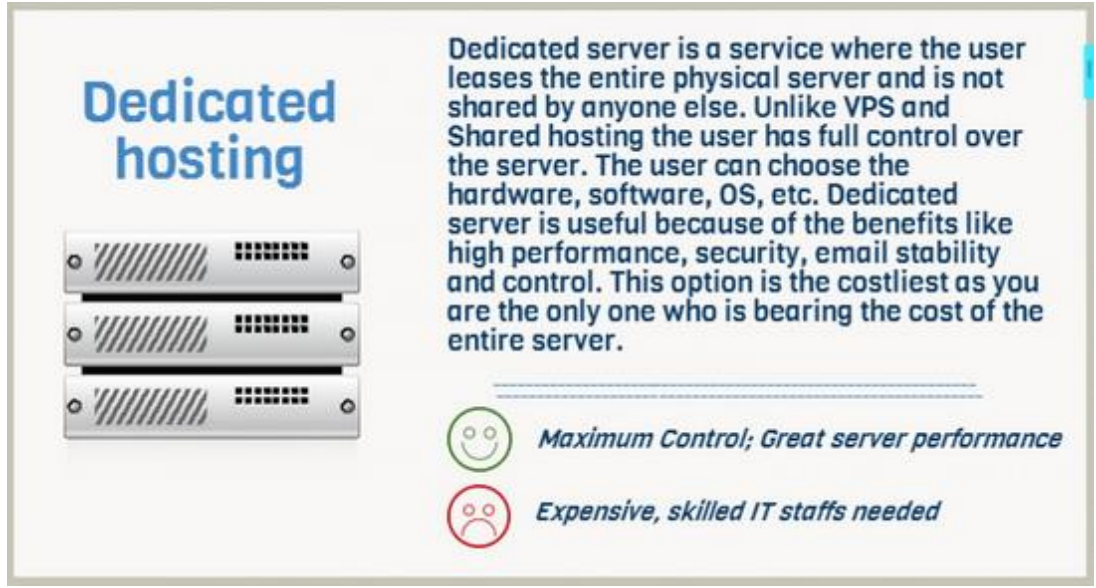
Figure 9: Dedicated hosting (Source: Visually & Rockcontent)

| Advantages | Disadvantages |
|---|---|
| <ul><li>Allows to use the server configuration and its CPU / RAM resources at 100%;</li><li>Fully customizable;</li><li>Cheaper than a virtual server.</li></ul> | <ul><li>More expensive than shared hosting,</li><li>Requires a technician to configure and maintain the server, especially for backups, security updates, RAID monitoring,</li><li>Non-redundant and non-resizable: risks of physical crash more and more important over the life of the server (hard disks, electronic cards, etc.),</li><li>It is generally necessary to change the dedicated server every 3-5 years.</li></ul> |

▷ **Virtual Private Server (VPS) hosting: the mixed solution**

A virtual private server is between the shared and the dedicated. It consists in making several virtual servers coexist on the same physical server, using virtualization software technologies. Thus, the hard disk of the physical server is shared between several virtual servers, each of which has its own central processing unit and its own memory allocation. The VPS thus allows more flexible administration, settings and installations than a shared server.

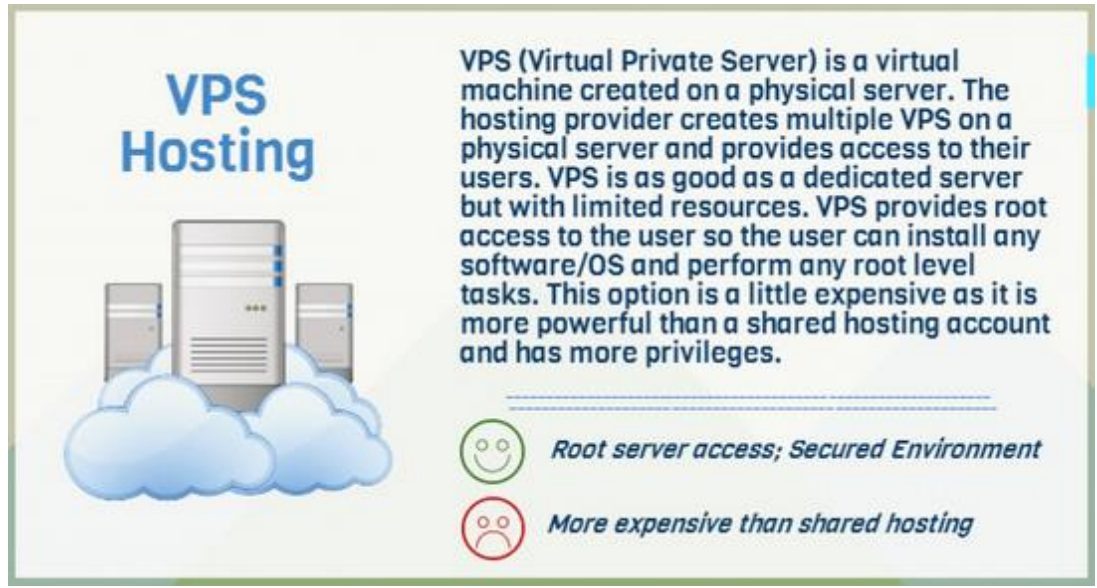
Figure 10: VPS hosting (Source: Visually & Rockcontent)

| Advantages | Disadvantages |
|---|---|
| - Easy installation, deployment and migration from one physical server to another;<br>- Potential gain on license costs (supported by the physical server);<br>- Simplified backups (snapshot of the virtual machine and its data). | - Dependence on a host server (in the event of a crash);<br>- Specific administrative constraints (security, backup);<br>- Little exchange or cooperation between IT administrators and database administrators. |

▷ **Cloud hosting: the flexible solution**

Cloud hosting is based on cloud computing technologies that allow an unlimited number of machines to act as a single system. Cloud hosting is not simply based on a physical machine divided into several virtual machines: it is based on the interconnection of several of these physical machines, itself divided into virtual machines.

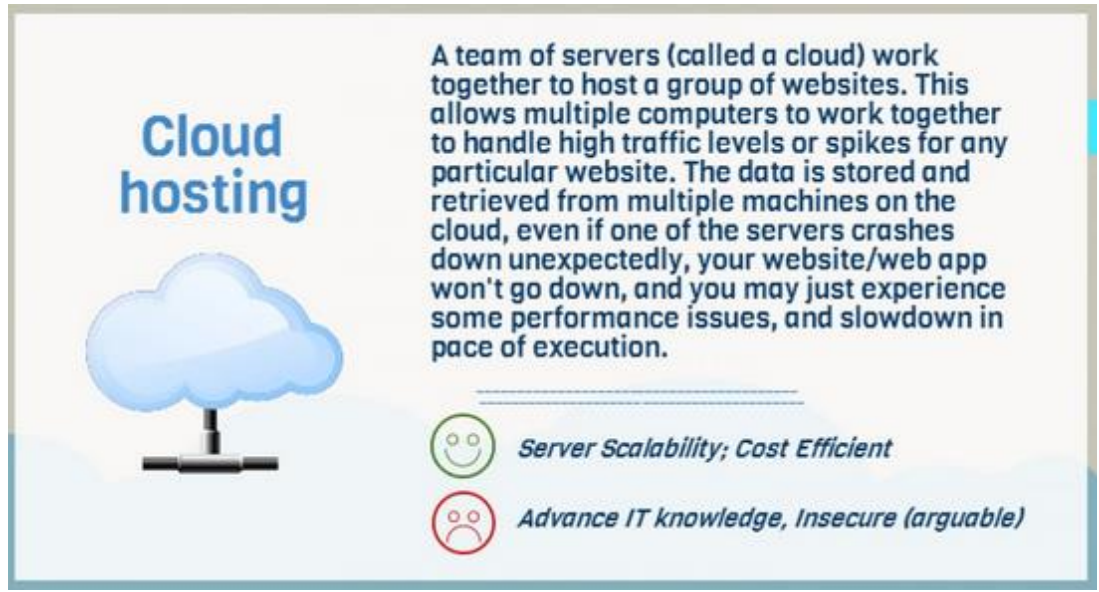
Figure 11: Cloud hosting (Source: Visually & Rockcontent)

| Advantages | Disadvantages |
|---|---|
| <ul><li>Guaranteed server resources (CPU, RAM)</li><li>Resizable data storage;</li><li>Flexible site growth;</li><li>Reduced prices and advanced features;</li><li>A potentially infinite lifespan.</li></ul> | <ul><li>Requires a technician to configure and maintain the server;</li><li>Dependent on the Cloud service provider;</li><li>Less flexibility than physical servers;</li><li>The most expensive solution.</li></ul> |

### 8.2.2 Hosting alternatives for an online course

It is therefore important to understand the challenges of the various site hosting solutions, whether they are shared, on a dedicated server, virtual or private, or in the clouds. Above all, we must understand the meaning of the indicators of the quality of services, performance, security, personalization (adaptive*) or even the flexibility that characterize one solution or another.

Hosting can also be done internally on the host institution's own servers if it has one. This choice would possibly involve the purchase of a server, but also skills for configuration and maintenance, unless the institution integrates the online learning system into its management policy for its existing computer system.

In general, in the absence of a previously known solution, it is recommended to take into account the factors of traffic density (low, medium or heavy) in order to opt for shared or dedicated hosting. The budget is also an important criterion even if the performance of the platform must be given priority. The key is to make sure you have good hosting and a website for several years, rather than having to migrate that site multiple times with probable high additional migration costs.

Ultimately, the hosting of Moodle remains dependent on several factors. Three plausible options for a university institution remain:

1. Hosting on the institution's server;
2. Hosting on the server of an external national or international operator;
3. Hosting on a server in the Cloud.

The choice of one solution or another normally takes into account the conditions of the host institution. It can be decided by applying the diagnostic form proposed in **Appendix 05.**

## 8.3 PREFERENCES IN IMPLEMENTING & CONFIGURING THE PLATFORM

The implementation and configuration of Moodle platform is well documented on the Internet. These two operations are done according to the type and the hosting support decided.

After the installing of Moodle and its configuration, a URL address, users and an administrator accounts are normally configured and given to the project leader or to a technical administrator of the platform. These access parameters allow the platform administrator to configure the course spaces, extensions (plugins) of the activities to be added, learners' registrations, etc.

This Guide is not intended to describe the steps of the server-side installing of Moodle or its technical configuration which remains dependent on the activities and extensions (plugins) that are selected to be installed.

However, this Guide points to potential tools for educational and communicative interactions that would help enrich an online learning environment. These tools make it possible to create, via the network, interactions between tutors*, learners and educational resources. Several types of plugins are to be updated from one version to another of Moodle.

### 8.3.1 Moodle plugins for online activities and resources

One of the great successes of Moodle, other than being free, is its ability to personalize online courses (adaptive learning*). A large number of plugins (extensions) are offered free of charge by Moodle collaborators who, once installed, can change the appearance and functionality of Moodle to adapt it to specific learning needs.

Few of plugins are installed by default the first time Moodle is installed. However, new extensions are permanently offered in an online directory of Moodle plugins[20].

A permanent review of installed extensions must therefore be done continuously according to the types of activities programmed in the online courses. This depends on the creativity and imagination of the online learning team, tutors* and teachers.

---

[20] Moodle plugins: https://docs.moodle.org/39/en/Resources

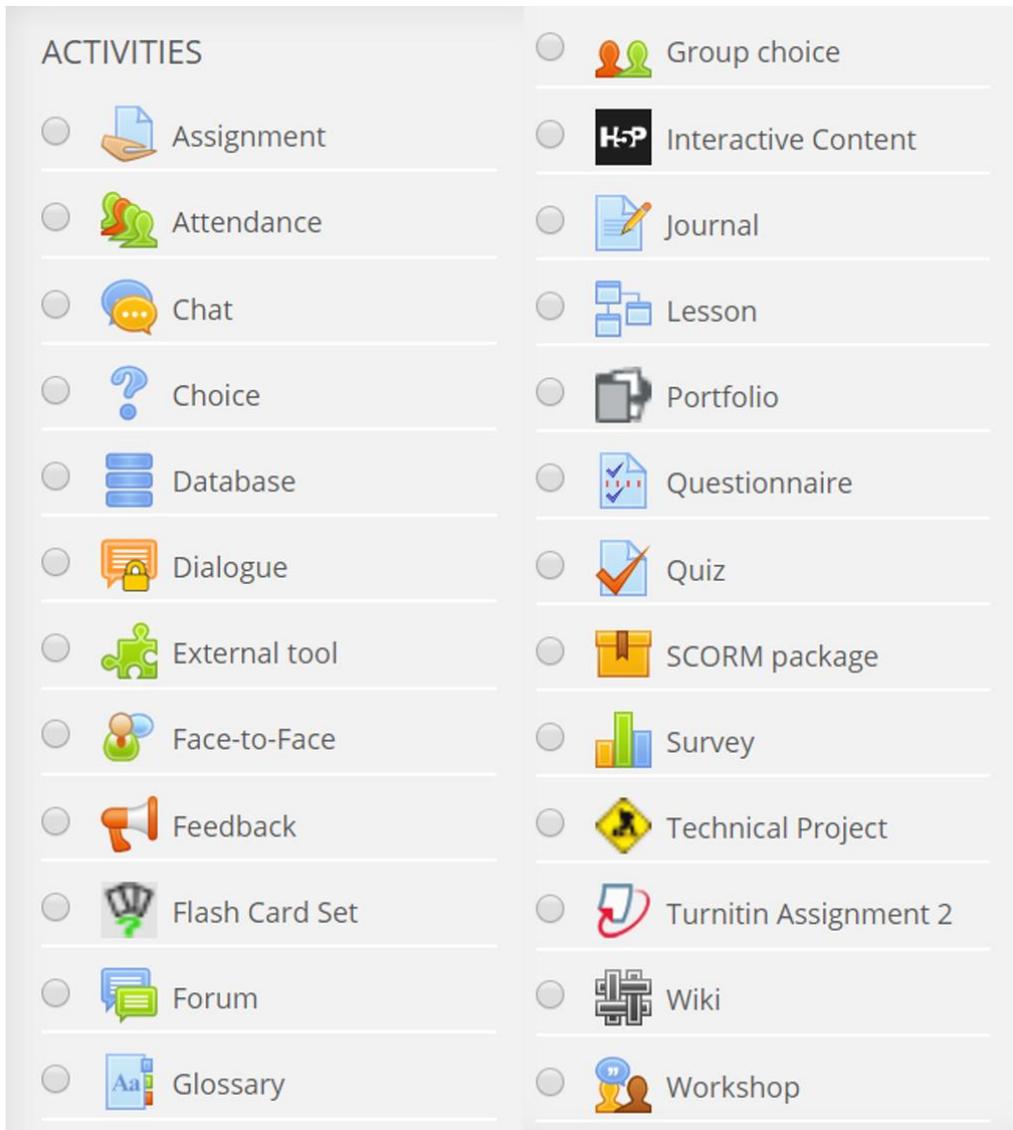

Figure 12: Examples of Moodle Activities and resources

The list of Moodle open source plugins is lengthy. It is listed in several categories (administration, assessment, collaboration, etc.) and several types (activities, notification, exercises and Quiz, etc.) for different versions of LMS Moodle.

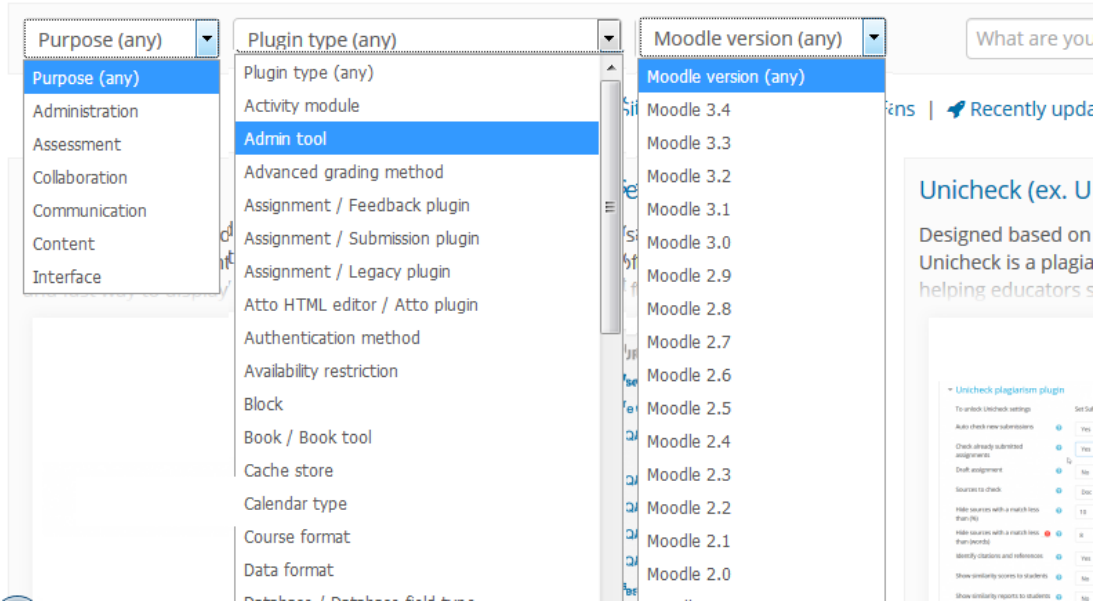
Figure 13: Additional plug-ins on moodle.org

In order to get a feel for it, here are some of the most popular and up to date ones to consider.

| | |
|---|---|
| 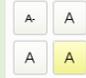 | **Accessibility Plugin**: The appearance of Moodle can be customized according to the individual preferences of each user. Customizations can include text size and colors. Changes can be made to all pages, and can also be saved permanently if desired. |
| 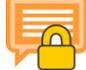 | **Dialogue**: This module allows students or teachers to launch two-way dialogues with another person. Although the official line is that this feature has been replaced by the messaging system, some users still find it useful. |
| 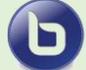 | **BigBlueButton**: This integrates BigBlueButton, an open-source web conferencing platform designed to provide a better learning experience for learners in a distance learning context. Learners can view slides and videos, and access Chat features via this plugin. It also records all sessions, so that learners can access them later. |
| 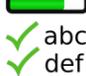 | **Checklist:** A great learner management tool, this plugin allows administrators or teachers of Moodle to create checklists for students, allowing them to keep track of what they have done and what they still need to complete. |
| 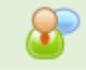 | **Face-to-Face:** Used to track, manage and remind learners of face-to-face training sessions that require advanced booking. Each session can be offered in one or more sessions. |
| 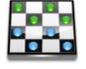 | **Game:** allows gamification to be integrated into online learning and training programs. We can integrate in questions and Quiz games like Hangman, crosswords, Sudoku, and snakes and scales, in your questions and Quiz |

### 8.3.2 End user's terminals additional software and plugins

Although Moodle prepares multimedia files for playback in the browser, the actual playback of certain resources is ensured by different types of browser-specific plugins (or add-ons), mainly Adobe Flash, QuickTime, Windows Media Player and Real Player. If users' computers do not have these pieces of software, they will be prompted to install them. These software components are generally free, easy to install, and widely used. Therefore, this would not be a major difficulty for most users.

Here are the download links for the most common software for upgrading the computer environments of end-user systems:

- Adobe flash player: https://get.adobe.com/fr/flashplayer/
- QuickTime: https://support.apple.com/fr_FR/downloads/quicktime
- Windows Media Player: https://support.microsoft.com/fr-fr/help/14209/get-windows-media-player
- Java and Java Runtime: https://www.java.com/fr/download/

# PHASE 02 - OPERATIONALIZATION OF THE ONLINE LEARNING SYSTEM

The second phase of supporting an online learning project consists in making operational the online system that was designed, approved, installed and configured during phase 01. This second phase constitutes the heart of the support operation, because it leads to the concrete implementation of the online learning system components.

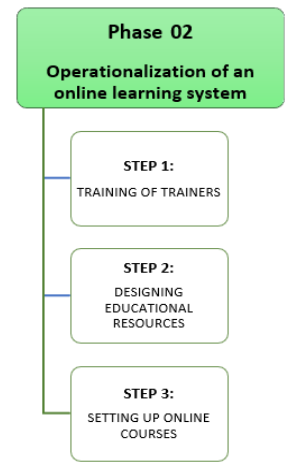

This phase is organized into three key steps:

1. Step 1: Training of trainers;
2. Step 2: Designing educational content resources;
3. Step 3: Setting-up online courses.

## 9   STEP 1: TRAINING OF TRAINERS

Before opening the Moodle platform to real use in a blended learning context, it is necessary first to train the entire project team on the educational activities and skills that will be necessary to make them autonomous after the support period. This underlies skills relating to the platform administration as well as the contents development, the tutorial accompaniment, learning scenarios scripting, knowledge assessment, and so many other innovative techniques and ICT-based educational activities.

However, at this level of support, training must follow a progressive chain in the acquisition of online learning skills. Initiatory training can be summarized in the following five topics:

1. Mastery of the general environment of an online learning platform, in this case Moodle;
2. Mastery of techniques for designing and structuring online courses;
3. Mastery of scenarios writing of learning pathways*;
4. Mastery of tutoring* functions;
5. Mastery of assessment modalities of learning outcomes.

These five fields constitute the core of a primary training program on the skills needed for an online learning. They can make the project team more autonomous to progress later to other forms of ICT based educational skills like flipped classes* and serious games.

Depending on the time provided for the training of trainers during this first period of support for the online learning project, training can be done according to 3 types of formulas:

1. "Packages" of training courses;
2. "On demand" training workshops;
3. "Compact" trainings.

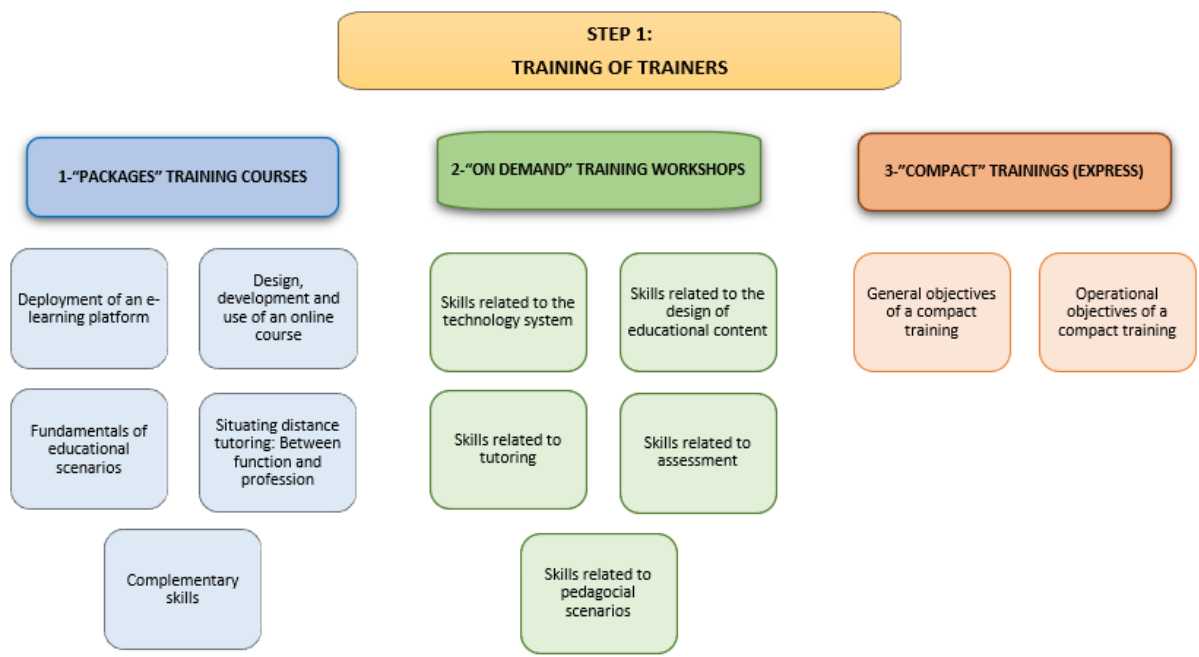

Figure 14: Flowchart of the training of trainers' three formats

To train the project team members on these founding skills, several repositories of training materials (Open Educational resources) are available online, including:

- The ICT training catalog of AUF/DRAP CNF/CNFp in Asia-Pacific[21];
- ICT UNESCO: a skills benchmark for teachers[22];
- Alison IT courses[23];
- Course Catalog: IT training and skills transformation[24];
- ERASMUS Training courses[25].

These catalogs are generally interoperable and provide complementary and shared training resources.

It is however likely to provide for a specific arrangement of tailor-made training linked to specific skills for a particular category of intervener in blended learning.

In all the packages (turnkey) training courses available online, some are well suited for initial training on the operationalization of an online learning, others are more oriented towards the appropriation of a specific aspect of the online learning.

In this stage of the training of trainers, we recommend the most important skills, in the following order of priority:

---

[21] Le catalogue des formations dans les CNF/CNFp en Asie-pacifique
[22] TIC Unesco: UNESCO ICT Competency Framework for Teachers: https://unesdoc.unesco.org/ark:/48223/pf0000213475
[23] Alison IT Courses: https://alison.com/fr/les-cours/it
[24] Course Catalog: https://www.globalknowledge.com/us-en/training/course-catalog/
[25] Erasmus training courses: https://www.erasmustrainingcourses.com/integrating-ict-into-teaching-and-training.html

1. Deployment of an "e-learning" platform: implementation, administration and integration of educational content. Example and practical case: Moodle;
2. Design, development and use of an online course;
3. Basic concepts of online learning and educational scenarios;
4. Situate online tutoring*: between function and profession;
5. Other additional trainings.

### 9.1.1 Course pathway "Deployment of an e-learning platform: Implementation, administration and integration of educational content. Example and practical case: Moodle"

▷ **Course objectives:**

"Following an introduction / presentation of the main functionalities of a pedagogical platform and of the different synchronous* and asynchronous* models, and of the open source world, the learner will be called upon to follow a series of pedagogical resources sequentially ordered step by step to deploy an online learning solution.

As a final module, an introduction to the integration and administration of the educational platform in a broader techno-educational environment will introduce the notion of "plugin", security, archiving, etc.

Learners will be called upon at the end of each module to do a collaborative work based on playing roles of installer, administrator, and editor of a course, so that at the end all learners access, as learners, the integrated course.

▷ **Teaching method:** face-to-face
▷ **Expected duration:** 03 days
▷ **Targeted skills:**

- Install and manage a massive and online learning platform [Master the process of installing a massive online learning platform, its administration and maintenance]

- Integrate educational content within a massive online learning platform [Master the basics of using a platform to upload educational content as well as the management of the main functionalities. To do this, the participant draws on his expertise as a teacher using face-to-face and distance learning, on the concepts presented during the training and on the tasks that he will have carried out on his own MOOC project]

### 9.1.2 Course pathway "Design, development and use of an online course"

▷ **Course objectives:**

At the end of the training, the skills acquired by the learner would allow him to:
- Know how to structure his course in order to put it in electronic format;

- Know the stages of development of an online course;
- Use a learning management system (Moodle);
- Master an authoring system of interactive exercises (Hot Potatoes);
- Define a scenario using his own online courses for training.

▷ **Teaching method:** face-to-face
▷ **Expected duration:** 03 days
▷ **Targeted skills:**

| |
|---|
| ▪ Produce teaching materials in digital format |
| ▪ Edit and share information on the Internet |

### 9.1.3 Course pathway "Fundamentals of educational scenarios"

▷ **Training objectives:**

"The training aims to master the fundamental concepts, models and tools of educational scenarios in order to:

- Design effective trainings by combining the targeted knowledge and the activities proposed to the learners in a structured and relevant way;
- Develop the learner's capacity to innovate by integrating new pedagogical approaches (the flipped class*, the project approach, serious games, etc.), new social practices (massive online courses, social networks, mobility), or new hardware technologies (e.g. augmented reality);
- Anticipate the operational implementation of scenarios designed for online learning platforms".

▷ **Teaching method:** face-to-face
▷ **Expected duration:** 03 days
▷ **Targeted skills:**

| |
|---|
| ▪ Design an e-learning system at each of its granularity* levels (learning spaces, lessons, activities, etc.); |
| ▪ Structure the learning system and design the learning scenarios; |
| ▪ Design scenarios for learner's accompaniment and support; |
| ▪ Design educational video sequences; |
| ▪ Use a techno-educational environment to implement the produced resources, activities and assessments in conformity with the pedagogical scenarios. |

### 9.1.4 Course pathway "Situating e-tutoring: Between function and profession"

▷ **Course objectives:**

Distance tutoring* (e-tutoring) is an essential component of online or blended learning courses. This training allows participants to build essential knowledge on e-tutoring method and principles. Its general objectives are to:

- Initiate a thinking on the practice of e-tutoring*;
- Develop distance mediator skills;
- Study the different e-tutoring modes according to learning situations;
- Share a common vocabulary on the concepts of tutoring* & e-tutor intervention methods;
- Vade-mecum of tutorial interventions and candidates' analysis of their own practices.

▷ **Teaching method:** face-to-face
▷ **Expected duration:** 03 days
▷ **Targeted skills:**

- Tutoring learners

### 9.1.5 Models of additional courses

In addition to these preliminary online learning courses, it is possible to consider other complementary courses (less difficult for this initiation phase) depending on the context (time and availability of learners and trainers). Among these complementary courses, it is possible to provide training on:

- Assessment in learning systems;
- Build and manage an online community;
- Produce an educational video;
- Reconsider the transmission of knowledge, skills and interpersonal skills by means of video games for educational purposes in a digital teaching context.
- Build the business model of an e-learning system.

## 9.2 ON-DEMAND TRAINING WORKSHOPS

If the previous workshop models do not meet the specific needs for training the trainers in this first initial stage of support for the blended learning project, or if their arrangements do not exactly correspond to the envisaged objectives of this step, "On demand" trainings can be adapted as required.

An on-demand training is made up of autonomous "skills", themselves made up of "knowledge", "know-how" and "Know-being". Any operator or sponsor of training can, as they wish, compose a learning content according to their own needs.

For the purposes of this first stage of training, this Guide offers a set of skills deemed useful for training the project team members to prepare them to their new pedagogical and technical environment of blended learning.

These skills are listed below according to a priority order as previously established, namely: "use of Moodle", "design of educational content", "scenarios", "tutoring*" and "assessment". They are available in "Knowledge", "Know-how" and "Know-being":

### 9.2.1 Skills related to the technological system (Moodle)

| Title of competence |
|---|
| ▪ Analyze the existing learning system, identify emerging needs and relevant solutions |
| ▪ Install a techno-educational environment taking into account the constraints related to the installation context |
| ▪ Define the general and specific objectives of an e-learning system |
| ▪ Install and manage a massive online learning platform |
| ▪ Configure a techno-educational environment taking into account the constraints related to the installation context |
| ▪ Integrate educational content into a massive online learning platform |

### 9.2.2 Skills related to the design of educational content

| Title of competence |
|---|
| ▪ Edit content on the Internet |
| ▪ Produce teaching materials in digital format |

### 9.2.3 Skills related to educational scenarios

| Title of competence |
|---|
| ▪ Build a learning progressive pathway from disciplinary contents |
| ▪ Design the scenario of learners' accompaniment and support |
| ▪ Produce educational video sequences |
| ▪ Produce a scenario of an e-learning course at each of its granularity* levels |
| ▪ Structure the learning system and design the learning scenario |
| ▪ Use a techno-educational environment to implement the resources, activities and assessments produced while respecting the educational and the support scenarios |

### 9.2.4 Skills related to tutoring

| Title of competence |
|---|
| ▪ Create and animate a community of practice |
| ▪ Develop simple note records to get started with tools and applications |
| ▪ Tutor and assist distance learners |

### 9.2.5 Skills related to assessment

| Title of competence |
|---|
| ▪ Design methods for assessing the learning outcomes |
| ▪ Develop an assessment strategy for the learning system |
| ▪ Assess the learning system with a view to improving its quality |
| ▪ Assess a techno-educational environment and its available tools in order to identify the most suitable for an eLearning system and eLearning activities |

## 9.3 "COMPACT" TRAINING (EXPRESS)

The third alternative for training the project team members can also be done in a condensed and quick manner. Instead of planning multiple training workshops around several separate skills and distributed over time according to differentiated learning pathways*, it is possible to set up a compact training plan that combines the priority skills described above.

The third scenario for training the project team in this initial stage can also be done in a condensed and quick manner. Instead of planning multiple training workshops around several separate skills and distributed over time according to differentiated learning paths*, it is possible to set up a suitable compact training plan that combines the priority skills described above.

### 9.3.1 General objectives of a compact training

A first type of compact training workshop aims to expose the team member to the wide-ranging use of information technology tools in an educational context. The main objectives are:

- Understand the contributions of different ICT tools to modernize its teaching practices;
- Acquire new skills on various ICT tools (editing, communication and assessment) and be able to integrate them in a face-to-face teaching as a starting point for an online course;
- Learn the basic functions of a distance learning platform before developing learning materials.

### 9.3.2 Operational objectives of compact training

▷ **Discovery training: Understanding the role of ICT for education:**

- Why using ICT skills for education (ICTE)?
- How to integrate the acquisition of these skills in learning?
- Are some digital skills essential?
- What skills are needed to integrate ICT in your classroom?
- What approach could improve the use of ICT in the classroom?

▷ **Initiation training: Learn to work on an LMS**

- Importance of educational platforms, panorama, selection criteria;

- Getting started with the main features of an educational platform (Moodle);
- Handling of tools allowing to work on the platform in student mode and in teacher mode;
- Introduction to educational scenarios (formulation of objectives, sequencing, implementation).

▷ **Advanced training: Assessment in an online course**

- Importance and accuracies of knowledge assessment in online learning;
- Know how to use an authoring system for interactive exercises, functionalities;
- Import/export Tests into/from Moodle;
- Know how to generate activity and performance reports on Moodle.

Each of these compact training models must provide project team members with a minimum of skills to migrate to a blended learning model. It is a priority to master the fundamentals of ICT based educational resources design and associated activities like tutoring* and assessment.

# 10 STEP 2: DESIGNING EDUCATIONAL RESOURCES

The periods devoted to resource design are an opportunity for project members to put into practice the skills acquired during one of the previous training sessions (Package, On-demand, or Compact trainings).

At this stage, the teachers of the project team are already trained on how to design and structure content according to the principles of online learning. They are relatively capable of performing the tasks assigned to them. However, in a starting phase, monitoring the quality of the resources to be produced is recommended. The quality of teaching and tutoring* will depend on it.

## 10.1 REMINDER OF THE PEDAGOGICAL DIVISIONS OF AN ONLINE COURSE

An educational resource (or medium) can take several names. The taxonomy is sometimes confusing between "syllabus", "online course", "educational object", "educational resources", "learning resources", etc. A course material also obeys a no less confusing subdivision between "part", "chapter", "module", "section", etc.

In educational jargon, the concept of "educational grain"* is also often used.

An educational grain* can correspond to a lesson or to a document in a lesson, etc.

The main thing in a structure as in another, is to make sure that the cuts made take into account the potential for autonomy of the cut entities (grains*) in order to be able to reuse them in different contexts. This is the principle of granularity*.

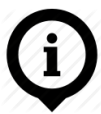

> By respecting the principles of granularity*, an educational grain* is the smallest semantic unit beyond which we can no longer cut out information. "Granularizing" a training medium means cutting its content into numerous items in order to be able to combine them again in different educational paths depending on the level and expectations of each learner. This allows the production of individualized learning pathways*.

## 10.2 REMINDER OF AN ONLINE COURSE CONSTITUENTS

By this general definition of a course structuring into modules, sequences, activities and resources, it is fundamental to remember that an online course is also composed of complementary resources (i.e. Text files, images, Video, URLs, Glossary, FAQ), learning activities (exercises), support (tutoring), communication (forum, chat), assessment (Quiz), etc.

It is therefore essential that the designer of a course combines a set of essential elements for the modules and activities on which the online course is mainly focused.

### 10.2.1 Elements of a module

- A definition of the prerequisites;
- A definition of general objectives;
- Identification of the targeted skills;
- A set of **sequences** containing activities.

### 10.2.2 Elements of a sequence

- Specific objectives;
- A summary plan;
- Scientific content;
- Learning **activities**;
- Additional resources;
- A form of formative assessment*;
- Interaction tools;
- A glossary;
- A list of frequently asked questions (FAQ).

### 10.2.3 Elements of an activity

- Instructions and expected results;
- Facilitation resources;
- Working methods (individual or / and collaborative);
- Assessment criteria;
- Interaction tools;
- Completion schedule;
- Self-assessments.

## 10.3 MONITORING THE DESIGN OF THE COURSE MODULES

The control and monitoring of these different stages, before the implementation of the online learning system, also makes it possible to identify the skills that need to be reinforced within the project team.

This monitoring and control work is done by regular collection of information from the project manager and content designers* through two documents:

1. A monitoring survey form used by the online learning project manager (**Appendix 06**);
2. A monitoring survey form used by the designers (**Appendix 07**).

These two grids summarize most of the recommendations proposed about what are called "design monitoring tools".

### 10.3.1 Monitoring grid by the blended learning project manager

The project manager is responsible for ensuring the smooth running of all project operations, respecting deadlines, the attendance of team members and especially the regular and efficient evolution of the process of designing educational resources and putting them online. Often, online learning organizers did not carry out such tasks, which requires learning, supervision and monitoring during the implementation period of the various operations.

It is for these reasons that the project manager is required to devote a lot of time to assist his teammates in defining priorities, developing timetables, defining respective responsibilities and functions to prevent possible tensions that may arise within the team or in its relations with other actors outside the institution.

One of the tools developed for this purpose is the monitoring survey form completed by the project manager (**Appendix 06**). It covers the main administrative and technical (and possibly economic) operations for monitoring and controlling the smooth running of operations. It is therefore important to complete this form regularly in consultation with the project team in the event of any difficulties or ambiguities.

### 10.3.2 Monitoring grid by content designers

The course designer monitoring form (**Appendix 07**) is a natural extension of the training workshops attended by the project team. It covers the main requirements that designers must meet: formulating prerequisites and objectives, structuring in sequences, designing diversified learning activities, designing self-assessments, etc.

The grid allows the designer to remember his skills in workshops and to apply them scrupulously to his course. By taking note of the grids filled-in by each of the designers, the project manager is able to identify the problem areas and take the necessary measures to deal with them.

# 11 STEP 3: SETTING-UP ONLINE COURSES

The next step is to load the developed online learning resources on the Moodle platform. To do so, it is important to follow a logical structuring process.

## 11.1 REMINDER OF THE ORGANIZATION PATTERNS OF AN ONLINE COURSE

It is first necessary to appropriate the skills of design and structuring of learning resources (course contents, simulation objects, assessment materials, etc.), in short, all the scientific and pedagogical material planned by the host institution to organize the blended learning project.

A course designer is expected to divide the content of his resources into units according to the organizational pattern adopted. These units vary according to the selected pattern and can have various names: lesson, chapter, section, module, activity, etc.

To use the terminology often used in Moodle, we propose a first pattern that presumes structuring courses into modules, themselves divided into sequences which are made up of activities and resources (see Figure 15). The combination of these different components to constitute a course translates into an educational scenario or learning scenario.

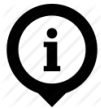
> A learning scenario represents the description made a priori (planned) or a posteriori (observed), of the course of a learning situation or learning unit aimed at the appropriation of a precise set of knowledge, specifying the roles, activities as well as knowledge manipulation resources, tools and services required to implement the activities. (J.P. Pernin, 2003).

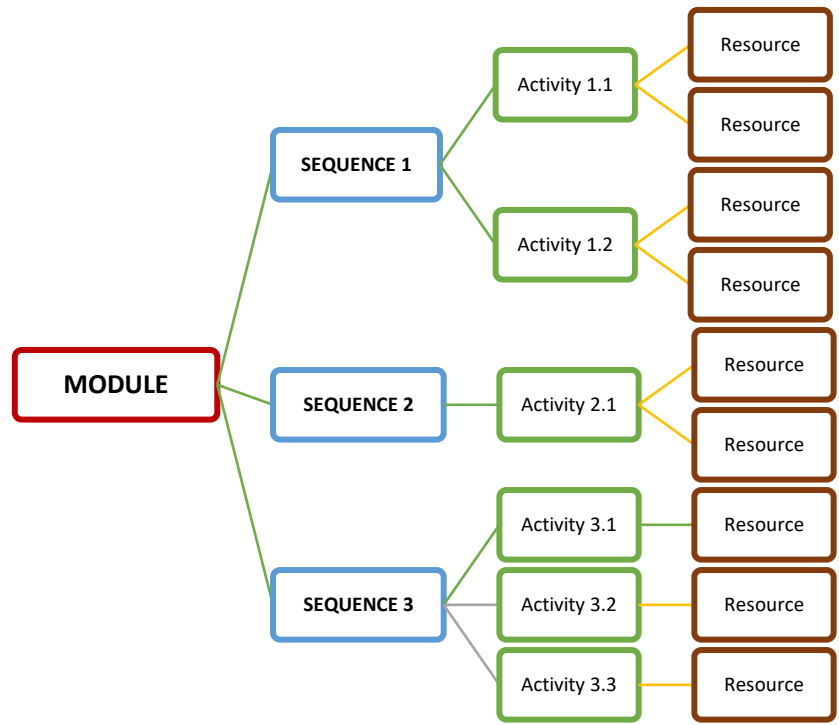

Figure 15: A module organization pattern (Moodle)

Several structuring patterns are possible. They depend on the density of the teaching material, the learning scenario, the original design of an existing course, etc. A sequence can, for example, correspond to a chapter just as it can correspond to a part or a section of a chapter. A single sequence can constitute a module, just as an activity can also constitute a sequence by itself. Sometimes a single resource in itself constitutes a sequence. In short, several factors come into account in an educational arrangement.

## 11.2 UPLOADING CONTENT RESOURCES ON THE PLATFORM

Courses or lessons are loaded on Moodle in spaces where teachers place learning materials for learners. In these spaces, the teachers deposit resources and learning activities which they organize according to the objectives and the teaching method to be followed in the learning sessions. For this reason, putting a course online could be, in a complex online course architecture, a process that might need several types of interventions from different contributors: content developer, tutor*, pedagogical engineer, platform administrator, learning manager, etc. However, in a simpler organization, a single person (tutor* or teacher), who has super user rights, can carry out on his own all the following operations:

- Adding online courses: creation of a new course space on Moodle platform;
- Modules creation: creation of one or more modules in the course space created for this purpose;
- Module organization: creation of module categories in a hierarchy useful for the different courses of an institution;
- Module configuration: control of contents displaying for learners;
- Adding module sequences if planned;
- Adding activities linked to sequences or directly to modules
- Adding resources into sequences;
- On Moodle, it is also possible to add content blocks to add additional data or plugins to a course such as Calendar, glossary, Blog, etc.

## 11.3 DEFINITION OF COURSE ACTIVITIES

As indicated before, a course is a set of contents and activities. Fourteen standard activities are defined on the Moodle site. They designate functions that the learner must perform alone, in interaction with other learners or with the tutor*.

- **Workshop** - allows peer review;
- **Database** - allows participants to create, maintain and search a bank of files;
- **Chat** - allows participants to have a synchronous chat in real time;
- **Consultation** - collects data from students to help teachers know their class and reflect-on their own teaching. Consultations are predefined (cannot be modified);
- **Homework** - allows teachers to grade and comment on files submitted by students, or an achievement made online or offline;
- **Feedback** - allows participants to create and carry out surveys in order to collect comments;
- **Forum** - allows to have asynchronous* discussions;

- **Glossary** - allows participants to create and maintain a list of definitions, such as a dictionary;
- **Lesson** - allows participants to deliver content in a flexible way, following different programmable pathways;
- **External tool** - allows participants to interact with LTI-compliant learning resources and activities on other websites;
- **SCORM** package - allows participants to integrate SCORM* packages into course content;
- **Survey** - allows a teacher to ask a question and to give a choice of multiple answers;
- **Test** - allows the teacher to design and include tests (quizzes), which can incorporate correct answers and / or automatic feedbacks;
- **Wiki** - a collection of web pages that anyone can create or modify alone or collaboratively.

These activities are defined according to a learning scenario established by the course designer and added to Moodle by the designer himself, the tutor* or any other person responsible for managing the online course. Often, the tutor* manages the moment of opening activities to the learners according to a scenario and a pre-established calendar.

Technically, a course material about the Moodle platform should explain these operations and provide simulations for setting up an online course.

# PHASE 03 - EXPERIMENTATION AND VALIDATION OF THE ONLINE LEARNING SYSTEM

The third phase of supporting an online learning project can start once the following conditions are fulfilled:

- the training of the project team members is carried out,
- the design process of resources and activities is completed,
- the assessment and self-assessment methods are defined,
- the general scenario of the project is completed.

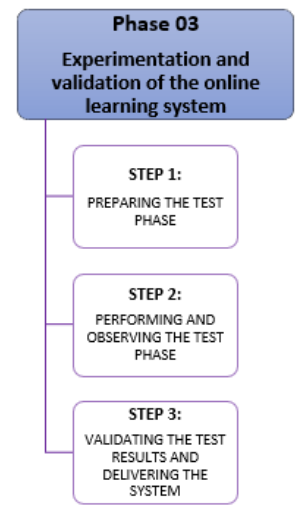

This phase includes an online operability tests of the learning system in the form of a short course simulation. The outcomes of this simulation constitute a key milestone in the validation process of the online learning system. If positive, the project is validated and an effective implementation of the online learning system can start.

This third phase is organized into three steps:

1. Step 1: preparing the test phase;
2. Step 2: performing and observing the test phase;
3. Step 3: validating the test results and delivering the system.

## 12 STEP 1: PREPARING THE TEST PHASE

The test phase must be carried out on a sample of volunteers chosen from within a real population of learners (students, teachers, administration staff). A team of teachers, trainers or tutors* need also be made up among the project pedagogical team. After experiencing the role of learner, they become more aware, from a learner side perspective, of the real conditions under which the future online educational environment will operate.

This test step is divided into five key moments:

1. Choosing a test object;
2. Recruiting volunteer students to participate to the testbed of the system (a test-group);
3. Appointing tutors* to supervise the test group;
4. Developing assessment questionnaires;
5. Test planning.

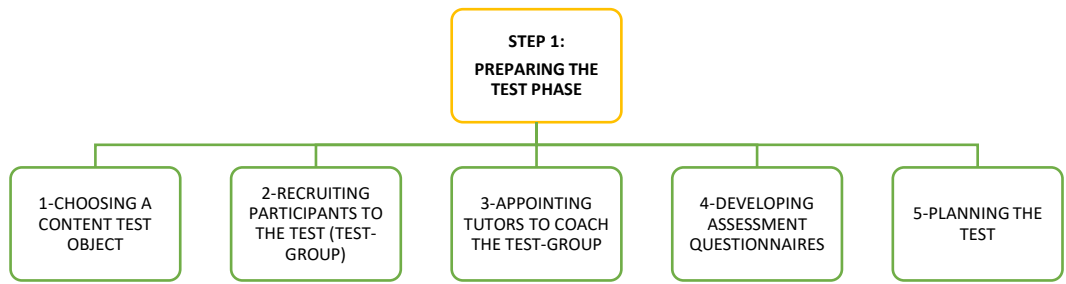

Figure 16: Flowchart of the Blended learning Test PHASE

## 12.1 Choosing a content test object

It is recommended to choose a module content having an average time length, comprising most of the characteristics to be tested: resources of various formats (text, audio, video, applications…), different educational approaches, bimodal interaction tools (synchronous* & asynchronous*), different working methods (collaborative & individual tasks).

The organization of this test module content would be better if it is structured in several sequences made up of as varied activities and resources as possible.

This module content can be planned in advance when creating online courses in step 02 of the precedent phase 02. It is even recommended to provide a transversal module content without taking into account the level of education. The idea is to have a heterogeneous population between several levels of university education. The main purpose of the test remains essentially the control of the proper technical and procedural functioning of the system more than a real transfer of knowledge.

## 12.2 Recruiting participants to the test (test-group)

Participants are selected with different profiles (geographic locations, usage of fixed and mobile terminals, different ages and levels of education). The test starts with putting them all in context during a face-to-face short meeting. The main goal is to explain to them the modalities of the simulation including how the platform works and how they should use it at distance. This quick training can be consolidated (or substituted) by training materials accessible on the institution's server in the form of tutorials, guides or video sequences.

In a real situation of a blended learning, short training sessions must imperatively be provided initially for learners to prepare them on how to use the system.

## 12.3 Appointing tutors to coach the test-group

The tutors* are chosen among the project team members. Some may never have done remote tutoring* before but they can be assisted by more experienced colleagues from the project team.

It is recommended to involve the largest number of tutoring* participants because they constitute the largest mass of actors in an online learning system after the learners. They outnumber content designers*.

It is therefore recommended to provide periodic tutors' training sessions throughout the life of the blended learning cycle. This type of training maintains a level of mastery of the online learning system which allows the teachers and tutors* to have a more precise idea of the concrete tasks of tutoring* and pedagogical support in a situation of distance education. Just as they will gradually be led to accept the principle of being assessed by their learners.

## 12.4 DEVELOPING ASSESSMENT QUESTIONNAIRES

Assessment is essential in an online learning. It is often classified into three types:

- Predictive assessment (or diagnostic) to know if the student could follow the training;
- Formative assessment to know if learning is progressing as expected;
- Summative assessment (or certification) to know if the objectives of the learning are reached.

But the most important principle of assessment is the fact to be carried out in the opposite direction: i.e. an assessment by the learners who give their opinions on the learning process and the tutoring* performance. This practice is not very common in traditional educational systems. It is even frowned upon in certain cultures or for certain teachers. However, it is more and more practiced as a means of revising and upgrading the teachings and curricula provided both face-to-face and at distance.

At this stage of the testing step, it is important to focus on the performance of the online learning system. The assessment process, targeting the learner's knowledge acquisition, is normally an integral part of the content of a course or a module. It is carried out by the content designer* and provided by the course tutor*.

On the other hand, the assessment of the training can be established by the training administrator or the educational manager to study the conditions under which the training takes place. In a way, it is a general assessment of the teacher's pedagogical performance and the general conditions under which the training takes place, seen and commented on by testimonials from learners.

Several assessment activities can be programmed on Moodle. These are either preconceived questions ("Survey" module) or questions defined from time to time by the course designer or tutor* ("Test", "Questionnaire").

Here following are some activities helping the assessments process.

▷ **Survey**

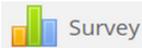 The "Survey" is a kind of predefined referendum in which the questions proposed are predefined and cannot be modified. Teachers can use them to collect data that will inform them of the students' assessments of the course nature and conditions (**Appendix 08**).

Teachers or tutors wishing to create their own questions will rather use the feedback activity (**Appendix 10**).

▷ **The questionnaire**

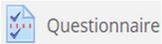 The Questionnaire is defined on Moodle website as an opinion-polling activity intended for the purpose of collecting data from users. This module allows the teacher to create a whole range of questions, for example to collect students' opinions about a course or activity. The Questionnaire module allows teachers to create a wide range of questions to get student feedback e.g. on a course or activities. The goals of the Questionnaire module are quite different from those of the Moodle "Lesson" or "Quiz" modules. With Questionnaire teachers do not test or assess the student, they gather data (**Appendix 09**).

▷ **Feedback or surveys**

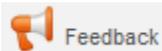 The activities of "feedback" type, also called "Surveys", can also be used by the teacher / tutor* to create a personalized survey questionnaire in order to collect information on the conditions of the course.

The Feedback activity permits to create and conduct surveys to collect feedback. Unlike the Survey tool, it allows to write one's own questions, rather than choose from a list of pre-written questions and unlike the Quiz tool, one can create non-graded questions. The Feedback activity is ideal for course or teacher assessments in order to improve the content for future participants. It allows anonymous surveys on course choices, school regulations, etc.

To develop this type of questionnaire for the Test period, you must refer to the Moodle documentation and also plan it as a training theme in the precedent phase 02.

Other assessment activities can also be planned to develop grids to be used either by tutors* to assess learners' progress or inversely by learners to assess tutors' performances and the course contents quality.

We introduce broadly some of these assessing activities but you need return to more detailed courses and guidelines on Moodle to discover their design techniques and ways of use.

▷ **Workshops**

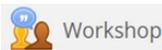 The workshop allows learners to assess themselves, which is likely to engage learners in a responsible collaborative activity increasingly sought after in the directions of active

pedagogy*. It is therefore recommended that course creators and teacher-tutors* become aware of the process (workflow) of this activity and supervise it closely (defining conditions) to avoid all forms of subjective assessment.

▷ **Tests and Quiz**

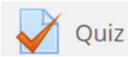  The activity module "test" allows the teacher to design and manage tests with questions of various types. Unlike the "Questionnaire" activity module, the test is more often used to assess the learner's knowledge.

▷ **Choice**

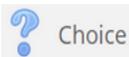 The Choice activity allows a teacher to ask a question and set up radio buttons which learners can click to make a selection from a number of possible responses. They can choose one or more option and they can update their selection if the teacher allows them. Choices can be useful as quick poll to stimulate thinking about a topic; to allow the class to vote on a direction for the course, or to gauge progress.

▷ **Hot potatoes quiz**

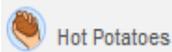 The Hotpot activity module allows teachers to administer Hot Potatoes and TexToys quizzes via Moodle. These quizzes are created on the teacher's computer and then uploaded to the Moodle course. After students have attempted the quizzes, a number of reports are available which show how individual questions were answered and some statistical trends in the scores.

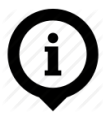 The control of these different modules and assessment activities must go through targeted trainings, programmed in the previous phase. Otherwise, dedicated guides and manuals are available on the Internet, particularly on the Moodle website[26].

## 12.5   PLANNING THE TEST

The project manager and the designated online learning coordinator are responsible for developing a test schedule. This schedule specifies the period, the tutors* involved, the dates and times of the synchronous meetings, the date of the informative* and summative* assessment, the dates of distribution and collection of the various questionnaires to be completed by the learners and the tutors*. It is based, among other things, on the learning scenario carried out when structuring the course or module.

---

[26] Cf. Moodle: activities. https://docs.moodle.org/39/en/Activities

# 13 STEP 2: PERFORMING AND OBSERVING THE TEST PHASE

The progression of the test phase must be done under conditions very similar to a real online learning situation. It essentially comprises the following moments:

1. An initial grouping face to face;
2. A tutored distant work;
3. Formative assessment activities;
4. Summative assessment activities;
5. Functional assessment activities.

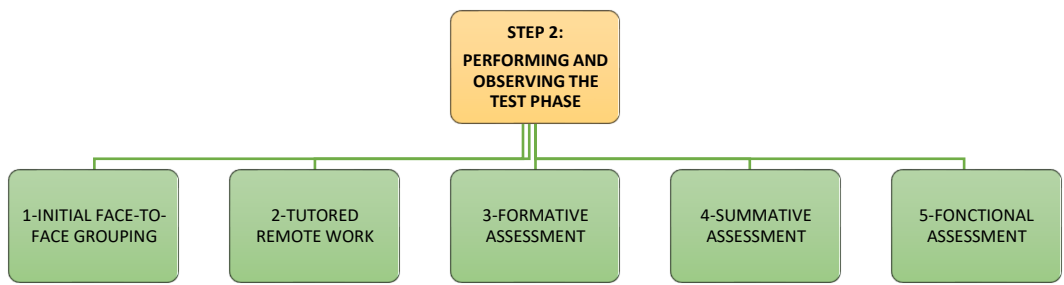

Figure 17: Organization of the Test phase of an online learning

## 13.1 INITIAL FACE-TO-FACE GROUPING

The purposes of this grouping (the organization and funding of which are the responsibility of the institution carrying out the project) are as follows:

- Establishing a first contact with the learners;
- Presenting the training module, the calendar and the working methods (learning scenario);
- Training learners on the use of the platform by insisting on the tools they will mostly use during their training: chat, instant messaging, forum, wiki, assignment submission, etc.

During this meeting, educational materials are distributed to participants: leaflets, manuals, programs, agenda, links to support resources.

Participants also receive their access parameters to the platform as learners.

## 13.2 TUTORED REMOTE WORK

During this stage, learners are guided by a schedule previously drawn up by the course designer and / or the tutor* who is / are at the origin of the learning scenario defined when setting up the course or module.

The schedule contains objectives to be achieved, activities to be carried out by learners and tutors* and a calendar of achievements and deliverables.

The schedule of tutored work must indicate in particular:

1. Synchronous activities and their dates;
2. Asynchronous activities, their durations and deadlines;
3. Formative and summative assessment* activities and their dates.

The tutored work stage continues throughout the entire test phase and must be punctuated by a "formative* and summative* assessment" period.

## 13.3 Formative assessment

The function of formative assessment* is to promote the progression of learning and to inform the student and the teacher about the skills or the elements to be improved. It targets specific learning and is part of one or more educational interventions. It is carried out during the activity and aims to report on the students' progress and enable them to understand the nature of their errors and the difficulties encountered. It can be led by the teacher, but can also be done in the form of self-assessment or peer feedback. No point, mark or percentage is associated with it.

This assessment is normally done in the form of a self-assessment by Quiz to allow the learner to know if he / she is correctly assimilating the concepts and concepts of the training.

Quizzes and quizzes should be set up to remain open to multiple tests, instant display of results, assistance and suggestion of clues to find the right answers, with no time limits or number of attempts.

The formative assessment* follows a logic of regulation, it aims to support the learning process, to help the learner to get closer to the learning objectives; it is therefore part of a helping relationship, a contract of trust, cooperative work.

The formative assessment* can be done on Moodle by several assessment activities including "Survey", "Workshop", "Test", "Questionnaire".

Teachers are required to carry out formative assessment* activities before proceeding to a summative assessment*. They must have offered students the opportunity to demonstrate their learning in a formative context allowing them to make mistakes, identify them and adjust themselves for the summative assessment*.

## 13.4 Summative assessment

Summative assessment is a form of acquired knowledge assessment. The purpose of summative* or certification assessment is to attest or recognize learning. It occurs at the end of a teaching process and serves to sanction or certify the degree of mastery of students' learning. It is the responsibility of the teacher and must be carried out in a fair and equitable manner, reflecting the achievements of the students.

This action is carried out at the end of the test phase by exploiting the potential of Moodle activities.

The difference with the formative assessment*, is that these summative assessments* (generally noted) must be configured to control the completion time, the uniqueness of the attempt, the non-postponement of the answer to the questions, etc.

## 13.5 FUNCTIONAL ASSESSMENT

The most important element in the test phase is the assessment made by learners and tutors* of the learning process.

Following the summative assessment*, learners and tutors* complete two standard forms, one for learners (**Appendix 09**) and the other for tutors* (**Appendix 10**). These two questionnaires allow everyone to express their opinions on the learning progression and to note its strengths and weaknesses.

These questionnaires are open to both profiles of tutors and learners at the end of the learning period. The responses are automatically analyzed and returned automatically by Moodle in the form of reports. These reports will be used by the project team to provide remedial solutions for areas of dysfunction that the two questionnaires have identified.

# 14 STEP 3: VALIDATING THE TEST RESULTS AND DELIVERING THE SYSTEM

Once completed and validated by the learners and tutors*, the reports are produced and edited and then transmitted to the project team (project manager, coordinator and/or pedagogical responsible) who will treat them both quantitatively and qualitatively.

The pedagogical college (project team) then meets to examine the results of this treatment and write the necessary recommendations concerning the solutions to be taken in relation to the forms of dysfunction recorded in the system before its validation and delivery.

Thus, this stage will be structured in 4 moments:

- System updating;
- Validation of project products;
- Final reports and the online system delivery.

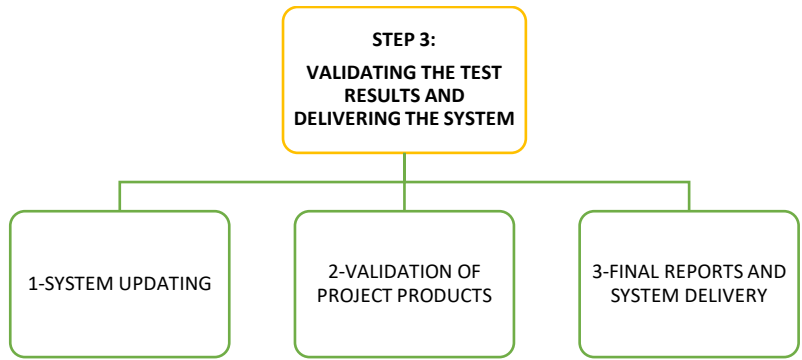
*Figure 16: Organization of the project validation stage*

## 14.1 SYSTEM UPDATING

The online system testing phase thus concludes with the implementation of the recommendations issued from the testing process. Detected malfunctions are corrected and integrated within the system. The project manager coordinates this operation and validates the corrections made to the various aspects of the system.

All the actors intervene, each according to his role and skills, on the points to be rectified: technicians, content designers*, tutors* and administrators.

The project requires an acceptable level of the online system operability during the blended learning period. The validation of the online system products needs a special care.

## 14.2 VALIDATION OF PROJECT PRODUCTS

The operability of the system is measured by the quality of its various aspects, including technical, administrative, educational and didactic. The economic aspect, even if it has not been dealt with in a very detailed precision in this Guide, due to the fact that no unique model can be applied, is set as a validation criterion which will impact the sustainability of the online learning system within the institution.

The indicators used to validate the project are to be recovered mainly from the results of questionnaires 09 and 10 (**Appendix 9 &10**) which indicate the satisfaction degrees of learners and tutors* on the progress of the Test phase.

Using learners and tutors' satisfactory results, the project team leader uses the form of **Appendix 11** to input his own appreciation criteria.

## 14.3 Closing report and project delivery

Based on the results obtained, a project closing report must be drawn up to validate the proper functioning of the online learning system before delivering it to be integrated into an official curriculum.

In this closing report, a set of resolutions are provided relating to the post-project period and to possible initiatives to enrich and mutualize this online learning experience.

After validation of the online learning system, the university institution now takes full responsibility for the system and ensures its supply, maintenance and dissemination to other courses or modules of its university curricula.

# CONCLUSION

In conclusion, it is important to note that any blended learning project constitutes a unique experience and that there is no only one packaged solution suitable for all situations.

This Guide is based on general rules and procedures. It should therefore be adapted to the reality of the context in which it is supposed to be applied. A specific "manual of procedures" should consequently complete this Guide in order to propose a tailored scenario for the operationalization of a concrete project. That manual of procedures should take into account the reality of the context of its application and serve as a dashboard for the blended learning project team.

Both the Guide and the Manual of procedures help to integrate a blended learning system within an existing face-to-face curriculum. Both can be readjusted during their application if readjustments come to be necessary.

The technical and procedural modalities in this Guide can also constitute the first elements of a MOOC learning system even if this massive online learning system entails a specific technical and pedagogical framework.

It is clear that the proposals in this Guide have been limited to basic educational activities and to most common monitoring and assessment practices. In a more sophisticated blended learning system, there would be more sophisticated techniques such as the integration of educational scenarios, flipped classes*, educational games (serious games), videoconferencing, and a whole range of more complex activities known under the label of "active pedagogy" *. These modalities would come in advanced phases after the implementation and appropriation of a first level online learning system. Possibly, the institution could later offer training on more advanced blended learning techniques and methods.

# BIBLIOGRAPHY

▷ **REFERENCE GUIDES & TUTORIALS**

- **Commonwealth of Learning** (2019). Blended Learning Practice.
  http://oasis.col.org/handle/11599/3218

- **OIF/IFEF.** (2016). OER trainer's guide v 1.1: Competency framework open educational resources—UNESCO Digital Library (p. 95).
  https://unesdoc.unesco.org/ark:/48223/pf0000266161

- **UNESCO**. (2011). Guidelines for open educational resources (OER) in higher education.
  https://unesdoc.unesco.org/ark:/48223/pf0000213605

- **UNESCO**. (2011). UNESCO ICT Competency Framework for Teachers—UNESCO.
  https://unesdoc.unesco.org/ark:/48223/pf0000213475

▷ **USED & RECOMMENDED READINGS**

- Alison IT Courses: https://alison.com/fr/les-cours/it

- Bennett D. et.al. (2010). "For-Profit Higher Education: Growth, Innovation and Regulation". Center for College Affordability and Productivity.
  http://www.centerforcollegeaffordability.org/uploads/ForProfit_HigherEd.pdf

- Bộ Giáo dục và Đào tạo. (2016). Thông tư số 12/2016/TT-BGDĐT Quy định Ứng dụng công nghệ thông tin trong quản lý, tổ chức đào tạo qua mạng

- Cleveland-Innes, M., & Wilton, D. (2018). Guide to Blended Learning. Commonwealth of Learning (COL). http://oasis.col.org/handle/11599/3095

- Coulon A. & Ravailhe M. (2003). « FOAD: économie des dispositifs et calcul des coûts ».
  https://www.centre-inffo.fr/IMG/pdf/economie_et_calcul_des_couts_foad.pdf

- Coulon A. & Ravailhe M. (2003). « Les coûts de la formation ouverte et à distance : première analyse » [http://sup.ups-tlse.fr/documentation/docs/fich_118.pdf]

- Druhmann, C., & Hohenberg, G. (2014). Proof of economic viability of blended learning business model. In Facebook Mediated Interaction and Learning in Distance Learning at Makerere University (ERIC Clearinghouse, 2014, p. 70-78).

- Erasmus training courses: https://www.erasmustrainingcourses.com/integrating-ict-into-teaching-and-training.html

- Financement et mise en œuvre de la FOAD : Vade-mecum des bonnes pratiques. www.una-univ-bordeaux.fr/Download/News/Info/document/233.pdf

- Global Knowledge Course Catalog: https://www.globalknowledge.com/us-en/training/course-catalog/

- Hanover Research Center (2013). "Business Models for Online Higher Education". https://www.hanoverresearch.com/media/Business-Models-for-Online-Higher-Education-1.pdf

- Loon, M. (2017). Designing and Developing Digital and Blended Learning Solutions. Kogan Page Publishers. https://books.google.fr/books?id=u141DwAAQBAJ

- Marc Weisser, « Dispositif didactique ? Dispositif pédagogique ? Situations d'apprentissage ! », Questions Vives [En ligne], Vol.4 n°13 | 2010, mis en ligne le 26 janvier 2011, consulté le 12 mars 2017. http://questionsvives.revues.org/271

- McGEE P. & REIS A. "Blended Course Design: A Synthesis of Best Practices". Journal of Asynchronous Learning Networks, 16(4), 7-22. 2012

- NGUYEN TAN Dai (2017), « Les TIC au service de la qualité des formations : le cas des programmes vietnamiens évalués par l'ASEAN University Network ». Thèse de doctorat, Université de Strasbourg

- Patrick, S., & Sturgis, C. (s. d.). Maximizing Competency Education and Blended Learning: Insights from Experts (p. 48). https://aurora-institute.org/wp-content/uploads/CompetencyWorks-Maximizing-Competency-Education-and-Blended-Learning.pdf

- PERAYA D. et. al. « Typologie des dispositifs de formation hybrides : configurations et métaphores ». AIPU. Quelle université pour demain ? Mai 2012, Canada. pp.147-155, 2012

- Powell, A., Rabbitt, B., & Kennedy, K. (s. d.). INACOL Blended Learning Teacher Competency Framework (p. 21).

- Quy chế về ĐT từ xa Ban hành kèm theo Thông tư số 10/2017/TT-BGDĐT ngày 28 tháng 4 năm 2017 của Bộ trưởng Bộ Giáo dục và Đào tạo

- Team, C. A. E. (2016). What is Blended Learning Method and how it works? CAE Computer Aided E-Learning. https://www.cae.net/blended-learning-introduction/

# GLOSSARY

- **ACTION PLAN**: An action plan favors the most important initiatives to meet certain objectives. In online learning, an action plan is constituted as a kind of roadmap offering a framework or structure of when it is necessary to conduct a blended training project.

- **ACTIVE PEDAGOGY**: It is a teaching approach which aims to convert trainees to actors of their own learning. Learners interact with each other and with the tutor*. The latter solicits them through synchronous and asynchronous exchanges. Participants are autonomous on certain sequences of the training and accompanied by the tutor* on other sequences.

- **ADAPTIVE LEARNING**: This term designates the idea of personalized teaching. The lessons, exercises, methods, adapt to each individual in real time, according to his learning pace.

- **ASYNCHRONOUS LEARNING**: In asynchronous learning, exchanges are not made live, but through discussion forums, or email, not allowing an immediate response.

- **BLENDED LEARNING**: Blended Learning designates a mixed learning mode: it is no longer a question of separating face-to-face learning and distance learning, but on the contrary of mixing them. The learner alternates between the two. He can for example start his learning with a training module at a distance, then put it into practice with a learning module in the presence of the trainer, and so on.

- **CONSTRUCTIVISM**: Constructivism, considers learning as a process of construction of knowledge which takes place in the interaction between the thinking subject and the environment in which it evolves. These theses grant an essential role to the actions and operations carried out by the subject in the structuring of thought. To build his knowledge, the individual uses previous knowledge as a means of representation, calculation and reflection on his own action. Old knowledge playing the role of new knowledge assimilation process. In other words, what an individual learns depends on what he already knows.

- **CONTENT DESIGNER** (of a course): Resource person, generally teacher or format (specialist), specialist in a subject or discipline, who develops knowledge content intended to be dispensed in a context of learning or learning according to objectives established within the framework of a learning offer.

- **EDUCATIONAL GRAIN**: Generic name to designate the smallest educational unit of an educational journey. The granule or the grain* is an educational object. Microscopic educational objects, elementary learning units are combined to constitute individual learning paths*.

- **EDUCATIONAL SUPPORT** (TUTORING): Support does not consist in transmitting knowledge. It is used to motivate the learner and to allow the learner to appropriate his learning path* at his own pace and according to his profile. A tutor* takes care of the trainee's problem and the answer in terms of a simple solution, within a fixed deadline. In an Open and Distance Learning system, we can distinguish a function of technical support, social support.

- **E-LEARNING**: The term e-learning designates distance education, and more specifically learning on the Internet.

- **FLIPPED CLASS**: Flipped or inverted classes designate a new educational approach, where classroom and home activities are reversed. Instead of learning lessons in class and practicing at home after school, students should learn the course at home, using suitable educational content, and will apply what they have learned in class. This method allows the teacher to be more available to learners in difficulty.

- **FORMATIVE EVALUATION**: It is practiced during the learning process and its aim is to inform the learner as completely and as precisely as possible about the distance which separates him from the objective to be achieved and about the difficulties he encounters.

- **GRANULARITY**: Level of cutting educational content into a series of elementary items, called grains*, which can be combined in the course of the course to meet individual learning needs.

- **INTEROPERABILITY**: Being interoperable is the ability of a product or a system, the interfaces of which are fully known, to work with other existing or future products or systems without restriction of access or implementation.

- **LCMS**: Learning Content Management System, are web solutions for creating and managing educational content. It is first of all a space where the educational contents are stored, in order to find them more easily. The space also includes tools for creating educational content for trainers.

- **LEARNING ACTIVITY**: Basic unit of a module for one or more specific objective (s) for a given skill. An activity groups together homogeneous and interrelated tasks which represent the most basic level of a given job (example: the tutor* performs several follow-up and support tasks included in his main tutoring* activity).

- **LEARNING MODULE**: unit integrated into a learning/training course. Element defined by objective (s), learning time, prerequisites, content, assessment. Composed of smaller educational grains*.

- **LEARNING PATHWAY**: This term designates all the learning modules to be followed by the learner. It meets several criteria: the specific need for learning and the level of competence held in the "taught" subject. It allows to individualize the course of the learning.

- **LEARNING SCENARIO**: Document presenting an ordered succession of teaching sequences staging human resources and multi-support which allow or contribute to the mastery by the learner of an objective and / or a professional skill. It is a planned organization of learning activities through educational resources, for a given educational objective and situation.

- **LMS**: Learning Management System: The term designates a web software system for creating online learning platforms, where groups of learners can be created and managed.

These platforms typically include a communication system, access control, and administration of learner groups.

- **MONITORING COMMITTEE:** A monitoring committee (MC) is made up of the project team de and the steering committee* to which are added the possible partners of the university who are associated with the online learning project (other universities, sponsors private sector, research structures, manufacturers, etc.). The main task of the monitoring committee is to watch over the good progress of the project towards its educational objectives. He intervenes when necessary on questions of a strategic order or relating to the governance model of the online learning system in place.

- **MOOC:** This is an English term for Massive Open Online Courses. These are online courses open to everyone, and often offered by universities. Participants, both trainers and students, communicate only via the Internet, and use free educational resources. Some MOOCs now offer certification at the end of the educational process.

- **ONLINE COURSE:** An online course is an educational course to be followed on a computer, tablet or smartphone, during which a learner will acquire knowledge and / or skills. It takes the form of a set of modules (which can be subdivided into sequences) which includes educational content and assessments. As the learner progresses, he can obtain certificates that validate learning.

- **ONLINE LEARNING SYSTEM:** Set of articulated elements (methods, tools, procedures, routines, principles of action) aimed at producing individual and collective skills; set of material and human means intended to facilitate a learning process.

- **PLATFORM ADMINISTRATOR:** IT device manager (manages accounts and permissions, general structure, course parameters, etc.)

- **PREDICTIVE EVALUATION:** Predictive assessment makes it possible to check the learners' prerequisites. It validates by tests or scenarios the prerequisites to integrate learning. She is therefore hired BEFORE training. We can call this a level audit, relevant to propose before an inter or intra-company session. The trainer can adapt his teaching scenario.

- **PRE-PROJECT:** In online learning, the pre-project is not the project specification. It is a first projection of what it could be in terms of the initial idea. It is necessary in a spirit of organization, to put the whole of the matter in perspective and to operationalize the approach. For each defined objective, the following questions will have to be answered: what actions? Which organization? What technical and educational, material and financial resources are necessary?

- **PROJECT TEAM:** The project team is internal to the institution hosting the blended learning project. It is made up of resource persons designated according to their profiles, roles and commitments in educational innovation by the online learning. The project team studies the needs of the university and prepares an application file to respond to the call for pre-projects of online learning. It is highly recommended that this team be versatile, made up of people with educational, technical, financial, legal and human resource management skills. Its main task is to prepare the most complete project file.

- **RESOURCE** (educational): A resource is composed of adequate and targeted content (aimed at responding to an identified need) and an adapted container ensuring media coverage (support, single or multiple, such as paper, magnetic support or support digital) which allow, in a learner-learner interaction (synchronous* or asynchronous*, face-to-face or distant), to implement an intention in the transmission, appropriation or sharing of knowledge or knowledge.

- **SCORM**: Sharable Content Object Reference Model. The term SCORM designates a coding standard for creating structured educational resources. It facilitates exchanges between the various online learning platforms.

- **SOCIO-CONSTRUCTIVISM**: An educational technique in which each learner is the agent of his learning and of group learning, by the reciprocal sharing of knowledge. The construction of knowledge, although personal, takes place in a social setting. The information is linked to the social environment and the context and comes both from what one thinks and from what others bring as interactions.

- **STEERING COMMITTEE**: A steering committee (SC) is formed as soon as the online learning project is validated. It is mainly composed of the university project team to which one or more external experts can be added. This mixed committee ensures the smooth running of the project according to the objectives and established operating procedures. The steering committee writes regular reports on the conditions of the activities during the project implementation period.

- **SUMMATIVE EVALUATION**: It is practiced at the end of an apprenticeship and its aim is to check if the objectives have been achieved by such or such learner.

- **SYNCHRONOUS LEARNING**: In synchronous learning, exchanges between participants, trainers and learners, are done using tools allowing a simultaneous conversation, such as a chat, videoconference, or web-conference.

- **TUTOR**: Accompany and help an individual or a group during the training. This assistance can be technical, moral, educational, administrative. This type of tutor does not intervene in the content (see teacher-tutor) but sometimes bridges the gap between teacher and learner.

- **TUTORING**: "Form of assistance in individualized teaching, which is offered either to support a learner who is experiencing difficulties, or to give specific, complementary or distance learning." "Tutoring can also be offered to small groups of individuals.

# APPENDICES

▷ APPENDIX 01: A pre-project presentation form: institutional framework

| PROJECT FORM ||
|---|---|
| Title of the pre-project: | Project code: |

| 1. PRESENTATION OF THE PROJECT (INCLUDE CONTEXT AND JUSTIFICATION OF THE PROJECT, OBJECTIVES, OVERALL STRATEGY AND EXPECTED RESULTS…) |
|---|

[one page maximum]

| 2. PROJECT LEADER |
|---|

University leading the blended learning project

- Name of institution:
- Name of the highest official of the institution:

Project leader of the blended learning within the institution
- Last name:
- First name:
- Position:
- Function:
- E-mail address:

| 3. PARTNERS IN THE PROJECT |
|---|

- Is the blended learning project under a cooperation agreement with

    - Members of the national socio-economic stakeholders?   Yes ☐   no ☐
    - Other university partners?   Yes ☐   no ☐

If yes:

- Lists of any other university partners

    - Name of the institution:
    - Contact within that institution:

    - Name of the institution:
    - Contact within that institution:

    - Name of the institution:
    - Contact within thar institution:

- **List of any socio-economic stakeholders**

    - Name of the stakeholder:
    - Contact within that stakeholder:

    - Name of the stakeholder:
    - Contact within that stakeholder:

    - Name of the stakeholder:
    - Contact within that stakeholder:

## 4. PROJECT CHARACTERISTICS

- **Domain and curricula of the blended learning project:**

- **Academic level of the proposed blended learning project:**

    - Bachelors: ☐
    - Master: ☐
    - PhD: ☐

- **Framework of the blended learning project**

    - National level: ☐
    - Offshore level: ☐
    - Double or joint diplomas: ☐

- **Status of the blended learning project:**

    - Initial learning: ☐
    - Professional training: ☐

- **Level of blending*:**

    - A full learning curriculum: ☐
    - A course in a learning curriculum: ☐
    - Module (s) in a course: ☐
    - Chapter(s) in a module: ☐
    - Other: ...........................................................................................
        *NB: Adapt taxonomy to the learning organization model in the institution*

- **Ratios between face-to-face and distance learning:**
    - Face-to-face teaching: …. %
    - Distance learning: … %

- **Distance teaching methods projected:**
    - Transmissive model [lectures]: Handouts, conferences, online courses, etc.: ☐
    - Activities [tutoring]: practical work, collaborative projects, remote support, etc.: ☐

- **Expected number of students enrolled in the proposed blended learning:** ……………………

- **Governance mode of the projected blended learning:**
    - A project committee has been created for the implementation and monitoring of the blended learning?

        Yes: ☐    No: ☐

- **If Yes, profiles of the project committee members:**
    - Administrative Officers:    ☐
    - Lecturers:    ☐
    - IT managers:    ☐
    - Other: ……………………………………………

- **Which of the following blended learning skills the institution defines proprietary for its project committee members?**
    - Learning platform administration:    ☐
    - Online course creation:    ☐
    - Educational scenarios:    ☐
    - Tutoring:    ☐
    - Active pedagogy:    ☐
    - Educational assessment:    ☐
    - Other: …………………………………………

- **Which of the following solutions the institution defines proprietary for preparing learners to blended learning?**
    - Intensive face-to-face training before the project starting:    ☐
    - Online free tutorials:    ☐
    - On-demand support according to the project need:    ☐
    - Other: ……………………………………………

- **The blended learning project is inspired by an existing blended learning system at national, regional or international level:**

    Yes ☐          No ☐

    If Yes, name some of them:

    - …………………………………………….
    - …………………………………………….

- **The institution possesses the following technologies in behalf of the blended learning project:**

    - Hosting server: ☐
    - Computer-equipped rooms: ☐
    - High Speed Internet Access: ☐
    - Wifi connection for students: ☐
    - Scanning equipment: ☐
    - Multimedia editing equipment: ☐
    - A videoconference system: ☐
    - Digital learning platform (CMS / LCMS): ☐
    - Others: ………………………………..

- **Language usage for the proposed blended learning:**

| Hourly volume per language | | | |
|---|---|---|---|
| National language | English | French | Other |
| Number of hours: | Number of hours: | Number of hours: | Number of hours: |
| % : | % : | % : | % : |

- **To what extent does the blended learning project bring social benefits for students (flexibility, mobility and time management, etc.)?**

    ……………………………………………………………………………………………………………………………………………………………………………
    ……………………………………………………………………………………………………………………………………………………………………………
    ……………………………………………………………………………………………………………………………………………………………………………

- **To what extent is blended learning an asset (an advantage) for the professional integration of students?**

    ……………………………………………………………………………………………………………………………………………………………………………
    ……………………………………………………………………………………………………………………………………………………………………………
    ……………………………………………………………………………………………………………………………………………………………………………
    ……………………………………………………………………………………………………………………………………………………………………………

- Will registration for blended learning be subject to additional registration fees for students?

  Yes ☐   No ☐

- The distance learning activities of the blended learning system will be:

  - Integrated into the global salary of teachers: ☐
  - Rewarded as additional activities: ☐

- The blended learning project will be allowed to use external services (tutoring, course digitization, Internet hosting, technical maintenance, etc.):

  Yes ☐   No ☐

- The financial resources for the blended learning project are:

  - Internal (from the institution): ☐
  - External (from partnerships): ☐

- The institution plans for an autonomous budget in behalf of the blended learning project:

  Yes ☐   No ☐

- The estimated ratio of financial support the institution requests from an external partner compared to the overall cost of the blended learning project:

  - 0-10%: ☐
  - 10-20%: ☐
  - 20-30%: ☐
  - 30-40%: ☐
  - More than 40%: ☐

- Describe internal elements to the institution that will contribute to the success of the blended learning project (human resources, material and financial resources, etc.):

  ..................................................................................................................................................................
  ..................................................................................................................................................................
  ..................................................................................................................................................................
  ..................................................................................................................................................................
  ..................................................................................................................................................................

- Describe how the institution could provide a positive Return on Investment (ROI) from the implementation of this blended learning project:

  ..................................................................................................................................................................
  ..................................................................................................................................................................

.................................................................................................................................................................................

.................................................................................................................................................................................

.................................................................................................................................................................................

- **What are / will be the difficulties and obstacles likely to be encountered, or already encountered, in the definition and implementation of the blended learning project? What measures are planned to overcome these difficulties?**

| Strengths | Weakness |
|---|---|
|  |  |

| Measures envisaged to overcome the identified difficulties |
|---|
| Depending on the identified weaknesses, the predicted measures can be presented as an activity developed within the framework of the blended learning project. The definition of each activity always begins with an action verb: strengthen, develop, structure, enable, etc. |

## 5. REQUEST FROM THE INSTITUTION CARRYING THE PROJECT

To be completed and signed by the highest manager of the institution carrying out the project or by the manager of its international relations department

Last name:
First name:
Title:

<div align="center">
Signature and stamp
(Required)
</div>

## 6. CONSTITUTION OF THE FILE

The pre-project file should include the following documents:
- **This application form,** duly completed;
- **List** of project committee **members** (For each member, indicate in the form of a table: surname and first name - Status (grade / function), Area of specialty, role in the project
- **Curriculum vitae** of the project leader;

All the documents indicated must be sent in **PDF** format.

An institution can submit several projects.

▷ APPENDIX 02: A pre-project presentation form: means, objectives & requirements

| PROJECT TITLE:<br>INSTITUTION:<br>PROJECT LEADER:<br>ACADEMIC YEAR: | |
|---|---|
| PRESENTATION ITEMS | X |
| **General project environment** | |
| The institution has already experience with blended learning systems | |
| The institution benefits from a national or regional technology support | |
| **Declared needs for the project** | |
| Need for managing a blended learning | |
| Needs for technical implementation support | |
| Needs for financial support | |
| Other needs | |
| **Declared objectives of the project** | |
| Clear educational objectives | |
| Clear scientific objectives | |
| Well defined objectives to improve the learning system | |
| Other: | |
| **Priority skills required to the blended learning project** | |
| Digital contents design skills | |
| Platform usage skills (Moodle) | |
| Learning support skills (tutoring) | |
| Writing educational scenarios skills | |
| Online assessment skills | |
| Other | |
| **Existing educational resources for the project** | |
| Digitized courses | |
| Exercise manuals | |
| Quizzes | |
| Other | |

| | | |
|---|---|---|
| **Technological infrastructure available favorable to the project** | | |
| | Accessible equipment to learners (PCs, tablets, etc.) | |
| | Local area network | |
| | Broadband Internet (ADSL) | |
| | Open connected spaces to learners (rooms, labs) | |
| | Videoconferencing system | |
| | Open Wifi access outside classrooms (common areas) | |
| | VPN network (access from outside) | |
| | Other | |
| **Profiles reported to be among the project team** | | |
| | Pedagogical Manager | |
| | Technical Manager | |
| | Educational resource designers | |
| | Tutor for learners | |
| | Other | |
| **Fund raising for the project** | | |
| | The project benefits from internal financial support | |
| | The project benefits from external financial support | |
| | **Total** | |

▷ **APPENDIX 03: Standard format of an opportunity memo of a blended learning pre-project**

## BLENDED LEARNING PROJECT OPPORTUNITY NOTE

**Project title:**

**INSTITUTION NAME:**

| | |
|---|---|
| Project leader: | |
| Domain / discipline / Theme: | |
| Academic level: | |
| Academic year: | |
| Training duration: | |

**I. QUALITATIVE EVALUATION OF THE PROJECT**

The abstract should not exceed 1,500 words using the following plan (2 pages maximum):

I.1 Quality of the technological framework of the institution (is the technical environment of the institution favorable to sustainable online learning empowerment?)

I.2 Quality of the project team (is the general level of the team sufficient to pool the skills acquired from the online learning to reproduce the experience alone?)

I.3 Quality of educational resources (are the existing educational resources sufficient to assess the quality of the planned online learning?)

I.4 Institutional anchoring of the project (to what degree does the project benefit from institutional support?)

I.5 Project contribution to the institution (can the project improve the quality of the university's learning curricula?)

I.6 Success factors (what could make this blended learning model a reproducible model?)

**II. JUSTIFICATION FOR AN EXTERNAL INTERVENTION**

(Advantages of the project for an external partner: how does the support of this project constitute an advantage for this partner?)

▷ **APPENDIX 04: A roadmap model**

**THE PROJECT NAME:**

(Indicative model to be adjusted to each situation)

**CONTEXT ELEMENTS**

- Project code:
- Host institution:
- Project title:
- Project leader:
- Referent:
- Academic year:

**PROJECT SYNOPSIS**

This paragraph summarizes the general objectives of the project.

**BLENDED EDUCATIONAL MODEL**

This paragraph determines the educational model of the blended learning system to be implemented: if necessary, identify one of the models of blended learning described previously ("blended learning typologies").

This section also fixes and prescribes the quotas of distance learning and face-to-face teaching in the blended learning project.

**METHODOLOGY: MANAGING THE DIGITAL REFERENTS OF THE PROJECT**

The institution designates a digital referent network for the project. This network covers learning, monitoring and assessment of activities. These tasks must be described and planned (who, when, how).

**STAKEHOLDERS, ROLES AND RESPONSIBILITIES**

*- The host institution:*

The host institution brings together persons in charge of controlling the smooth running of the project. These persons take strategic decisions. Indicate from which direction, service, department, sector, project, etc. the blended project to set up depends of.

*- The project team:*

The project team is headed by one and only one project manager. Specify the areas of expertise, roles, missions and responsibilities of each member of the team.

*- External contributors:*

If the project team does not bring together all the expertise necessary for the smooth running of the project, it is necessary at this stage to have identified external experts (tutors, resource developers, technicians, etc.) that the team can request for collaboration at the appropriate moment under appropriate conditions (including financial).

*- End users:*

Specify the category of end users of the system: academic level, number, single stream or pooling, etc.

DIGITAL ELEARNING ENVIRONMENT

*- The system definition:*

Define the technical configuration of the educational system: specify the choice of educational platform to use, decide on the place and conditions of its hosting, list all the tools, applications and additional services useful for learning.

*- Connectivity and remote access:*

Identify the solutions to be deployed to facilitate remote access to the platform and maintain virtual contact among the various eLearning players: specify the mode of access to the platform outside the institution (VPN network, tunneling, roaming, roaming profiles); indicate whether access is planned in adaptive mode (responsive design) via PC, Tablet and smartphone; specify whether videoconferencing solutions are planned and what type (via web interface or via specific IP protocol).

SETUP WORK

*- The conditions for producing educational content:*

Describe the choices and conditions under which the digitization of educational resources should be done. Define, among other things, the content types: courses, exercises, Quizzes, Educational videos, SCORM* aggregations, Web resources, etc. Also define the quality of the resources to be used: new content, existing content, innovative content, scripted content, etc. Indicate the choices in terms of rights and licenses: OER or / and proprietary resources.

*- The conditions for carrying out educational support (Tutoring):*

Identify the tutors and the conditions of their appointments. Among other things, define whether the tutors are internal to the institution, whether they are both course designers and tutors of their own resources, whether they are paid or not. Indicate also if there is an appeal to external contractual tutors, their academic levels and the rate of their involvement.

## TRAINING OF THE PROJECT TEAM

Define a precise learning schedule for the members of the project team (content designers*, trainers, tutors, technicians, administrators). As a general rule, everyone involved in the development of a blended learning must undergo training on the type of activities they are supposed to perform on the educational system.

## SCHEDULE MANAGEMENT

### - Director planning:

The project master schedule is a synthetic project schedule limited to the main guiding milestones (Submission of the project, validation, start-up, mid-term review, closure). It is one of the main tools of strategic management. Apply the most visual representation possible (timeline, charts, grids, concept maps).

### - Operational planning:

Operational planning is often carried out in the form of a Gantt chart which allows a project manager to monitor the progress of all the stages or milestones that make up the project. At this stage of the blended project, le project manager must detail the master schedule.

## COST MANAGEMENT (BUSINESS MODEL)

At this stage, the workloads (tasks, activities or deliverable) must be identified. These charges can be calculated in terms of costs according to the economic rules applied by the institution. Estimating the costs of these charges is the essential part of the economic model of the project.

## MONITORING AND REPORTING

### - Operational communication tools:

Define how communication will take place between the members of the project team and with the project manager: type of media (paper, email, collaborative application, etc.), type of content, frequency, etc.

### - The meetings:

Define the nature and mode of meetings between the different project teams. For each meeting: frequency, duration, place, host, who sets the agenda, who writes the minutes, who hosts, etc.

### - Reporting:

Define the nature and frequency of reports communicated by the steering committee* on the progress of the online learning project.

RISK MANAGEMENT

This is to make an inventory of events likely to disrupt the smooth running of the project and to assess the impact of these risks on the project. These risks can be of a technical, human or financial nature. They can be internal or external to the institution. However, there is no point in listing risks without providing for countermeasures to avoid them or limit their impact.

PROJECT CLOSURE AND REPORTS DELIVERY

The closure of the project is declared between the two parties by a set of documents which include in particular the final report of the project leader. This report must provide indicators which make it possible to understand the educational quality of the project: participation rate and attendance of participants, dropout rate, volume of work submitted, real rate of remote access and face-to-face, general success rate, etc.

The report must also summarize all of the real problems and difficulties encountered both at the technical level (platform breakdowns, computer network problems, electrical current problems, etc.) and human (absence of tutors at meetings, learners nonattendance in synchronous sessions, etc.) or financial (insufficient resources for certain activities such as tutoring*, development or acquisition of educational material, etc.).

▷ **APPENDIX 05: Technical framework of a blended learning project**

This diagnostic grid allows an online learning host institution, which has its own server, to determine whether it can host a Moodle-type platform and the traffic it generates during the progress of distance learning.

This grid can also be used to analyze the performance of external service proposed to host the technical system of the online learning.

1. **TYPE OF HOSTING SERVER**

Which of the following types does the institution's server correspond to?

  ☐ Shared server;
  ☐ Dedicated server;
  ☐ Virtual server;
  ☐ Server in the clouds.

2. **STORAGE (ALLOCATED DISK SPACE) ON THE SERVER**

The disk storage space determines the number and size of files that can be stored on the server or the server portion reserved to the eLearning platform.

  ▪ Mo = ...
  ▪ Go = ...

3. **THE SERVER DATABASES**

The database type of the system determines the techniques used to import and export data to and from the eLearning system.

| Type | Quantity | Size | Maximum number of simultaneous connections |
|---|---|---|---|
| MySQL | | | |
| Oracle | | | |
| SQL Server | | | |
| Other | | | |

4. **THE SERVER BANDWIDTH**

Bandwidth is the amount of data allocated by the host to display each element that makes up a site (html pages, images, videos, documents to download and even requests and responses from the database). The greater the weight of pages and images, the more bandwidth is required. As an indication: 100 MB of bandwidth represents 10,000 monthly visits with an average of 10K each. A bandwidth of 2 to 5 GB should normally be satisfactory.

  ▪ Downloading speed per second:
  ▪ Uploading speed per second:

## 5. SERVER DOMAINS AND SUB-DOMAINS

The number of sub-domains can be of great importance in the event that CDN (Content Delivery Network) needs arise. A CDN is a means of externalizing data or a section of its site in order to avoid overloads on the same server.

- Number of sub-domains of the main server: ......
- The learning project will have its own domain name:  ☐Yes        ☐No

## 6. THE SERVER CONFIGURATION

The two main families of operating systems used are Windows or Unix (or Linux). The web server is controlled by an HTTP server, the most popular and most used of which is the Apache server. An optimal configuration of the Apache server is particularly essential to allow the site to achieve good performance, especially in terms of page loading speed.

- Server OS: ❑ Linux ❑ Windows ❑ Other: ............
- Http server type: ❑ Apache ❑ Microsoft  ❑ Google Web Server ❑ Other: ............
- Programming languages: ❑ PHP ❑ ASP ❑ Other: ..................................

## 7. TECHNOLOGIES ACCEPTED BY THE SERVER

To ensure good server maintenance and efficiency over time, it is important to have a web server that manages the latest technologies.

- ☐ Php5.4;
- ☐ CGI scripts;
- ☐ Zend;
- ☐ Optimizer;
- ☐ Others: ..........................

## 8. MAIL MANAGEMENT

Two protocols (POP and IMAP) manage the e-mail service on a server. With IMAP all mail and message folders remain on the server and are therefore accessible from any computer connected to the Internet. POP retrieves mail from a remote machine when you are not permanently connected to the Internet (less and less recommended).

- ☐ IMAP (Interactive Mail Access Protocol)
- ☐ POP (Post Office Protocol)

## 9. SERVER SECURITY OPTIONS

Antivirus options, spam, blacklists are essential for the safety of the server.

- ☐ Antivirus;
- ☐ Antispam;
- ☐ Antiphishing;
- ☐ Blacklisting;
- ☐ Other: .....................

## 10. TECHNICAL SUPPORT

A good host must be easily reachable, either by phone or, at a minimum, by email. The responsiveness of the hosting team must largely weigh in the choice of web hosting.

- ☐ Hotline 24/7;
- ☐ Dedicated team;
- ☐ Online help;
- ☐ Other: ……………………

▷ **APPENDIX 06: Monitoring survey form by the project manager**

| |
|---|
| To be completed by (name/role): |
| Date: |
| Module code: |
| Title of the module: |
| Course title: |

<div align="center">**CONTENT DESIGN: THE DESIGNERS**</div>

▷ **Item n ° 01**

- The list of course content designers* is established and the work contracts (agreements) are validated:

  ☐ Yes
  ☐ No

The list of designers must be drawn up and validated from the start. This list help creating the designers' accounts on the Moodle platform.

▷ **Item n ° 02**

- All designers are informed of the roadmap and deadlines for finalizing the design work:

  ☐ Yes
  ☐ No

It is essential to consult with the whole team on the timetable and the roadmap and to obtain their support in order to avoid planning conflicts.

▷ **Item n ° 03**

- Content designers are informed of the availability of the Moodle platform and its URL address and their account settings for accessing it remotely:

  ☐ Yes
  ☐ No

▷ **Item n ° 04**

- Content designers have received their platform access settings and all have validated their designer accounts:

  ☐ Yes
  ☐ No

▷ **Item n° 05**

- The designers have assimilated the framework recommended for the design of the modules:

  ☐ Yes
  ☐ No

▷ **Item n° 06**

- For any case, indicate the measure (s) to be taken to ensure that all designers assimilate and comply with the recommended design framework:

  ……………………………………………………………………………………………………………………………………………………
  ……………………………………………………………………………………………………………………………………………………

▷ **Item n° 07**

- Describe the access problems to Moodle which were brought to your attention by the designers and the measures to be taken so that each designer can do so without problem:

| Problems encountered | Types of solutions proposed |
|---|---|
|  |  |
|  |  |
|  |  |
|  |  |
|  |  |

▷ **Item n° 08**

- Describe the problems in using functions Moodle that have been brought to your attention and the measures to be taken to allow each designer to use them properly:

| Problems encountered | Types of solutions proposed |
|---|---|
|  |  |
|  |  |
|  |  |
|  |  |
|  |  |

## CONTENT DESIGN: THE TEACHING TEAM

▷ **Item n° 09**

- All of the courses / modules to be designed for the online learning are final and the contents have been validated within the host institution (by an educational college)

  ☐ Yes
  ☐ No

The table of modules must be completed and validated by the pedagogical college of the university hosting the eLearning system. This list allows that permissions will be granted to designers to access course spaces on the learning platform.

▷ **Item n° 10**

- Scientific module managers have been appointed to validate the scientific contents produced by content designers*:

  ☐ Yes
  ☐ No

The educational college ensures the scientific quality and legitimacy of the designed modules. It gives recommendations for upgrading the modules and ensures the quality of assessment.

▷ **Item n° 11**

- The members of the pedagogical college are informed of the need to validate the modules contents as they become available:

  ☐ Yes
  ☐ No

▷ **Item n° 12**

- The exchange conditions between the designers and the pedagogical college have been established to ensure regular validation:

  ☐ Yes
  ☐ No

▷ **Item n° 13**

- An eLearning administrator has been appointed and his account is validated on the platform:

  ☐ Yes
  ☐ No

▷ **Item n° 14**

- What are the steps already carried out for the preparation of the blended learning promotion and management?

|  | In progress | Completed | Not started |
|---|---|---|---|
| ▪ Application call/form for future students | ☐ | ☐ | ☐ |
| ▪ Advertising brochure of the blended learning project | ☐ | ☐ | ☐ |
| ▪ Estimated budget | ☐ | ☐ | ☐ |
| ▪ Tutorial charter | ☐ | ☐ | ☐ |
| ▪ Tutor contract | ☐ | ☐ | ☐ |
| ▪ Designer contract | ☐ | ☐ | ☐ |
| ▪ Coordinator contract | ☐ | ☐ | ☐ |
| ▪ The platform template | ☐ | ☐ | ☐ |
| ▪ List of frequently asked questions | ☐ | ☐ | ☐ |
| ▪ Course agenda | ☐ | ☐ | ☐ |
| ▪ Exams calendar | ☐ | ☐ | ☐ |
| ▪ Knowledge assessment strategy | ☐ | ☐ | ☐ |
| ▪ Others | ☐ | ☐ | ☐ |

▷ **APPENDIX 07: Monitoring survey form by the content designer**

To be completed by (name/role):
Date:
Module code:
Title of the module:
Course title:

▷ **Item n ° 1**

- Have you formulated the objectives of your course clearly and precisely?

  ☐ Yes. I'm sure;
  ☐ Yes. I think so;
  ☐ No. I cannot define the objectives;
  ☐ No. I haven't had time yet.

▷ **Item n ° 2**

- Have you divided your course into learning sequences corresponding to the objectives that you have defined?

  ☐ Yes. Absolutely;
  ☐ Yes. But I am not yet satisfied with the breakdown;
  ☐ No. I can't do it alone;
  ☐ No. I haven't had time yet.

▷ **Item n ° 3**

- For each sequence, have you planned:

|  | Yes. For all sequences | Yes. Only for certain sequences | No. I find it hard to fulfill this requirement | No. I don't have time yet |
|---|---|---|---|---|
| A description recalling the objectives, the prerequisites | ☐ | ☐ | ☐ | ☐ |
| A graphic organizer | ☐ | ☐ | ☐ | ☐ |
| Course materials | ☐ | ☐ | ☐ | ☐ |
| Learning activities | ☐ | ☐ | ☐ | ☐ |
| Additional resources to consult | ☐ | ☐ | ☐ | ☐ |
| Formative assessments | ☐ | ☐ | ☐ | ☐ |
| Interaction Tools | ☐ | ☐ | ☐ | ☐ |
| A glossary of concepts | ☐ | ☐ | ☐ | ☐ |
| A list of frequently asked questions | ☐ | ☐ | ☐ | ☐ |

▷ **Item n ° 4**

- For each activity, have you planned:

| | Yes. For all activities | Yes. Only for certain activities | No. I find it hard to fulfill this requirement | No. I don't have time yet |
|---|---|---|---|---|
| A description recalling the work instructions, the expected result | ☐ | ☐ | ☐ | ☐ |
| Resources to facilitate the achievement of the activity | ☐ | ☐ | ☐ | ☐ |
| The working method (individual / collaborative) | ☐ | ☐ | ☐ | ☐ |
| The assessment criteria | ☐ | ☐ | ☐ | ☐ |
| Interaction Tools | ☐ | ☐ | ☐ | ☐ |
| A completion schedule | ☐ | ☐ | ☐ | ☐ |

▷ **Item n ° 5**

- Have you planned for each self-assessment activity?

| | Yes. For all self-assessments | Yes. Only for certain self-assessments | No. I find it hard to fulfill this requirement | No. I don't have time yet |
|---|---|---|---|---|
| Feedback(s) | ☐ | ☐ | ☐ | ☐ |
| Links to the concerned resources | ☐ | ☐ | ☐ | ☐ |
| A help system in case of difficulty | ☐ | ☐ | ☐ | ☐ |

▷ APPENDIX 08: Assessment by the learner: a pre-defined grid on Moodle

☐ Relevance

| Answers | Not yet answered | Almost never | Rarely | Sometimes | Often | Almost always |
|---|---|---|---|---|---|---|
| In this online course ... | | | | | | |
| 1. My learning focuses on subjects that interest me. | | | | | | |
| 2. What I learn is important for my professional practice. | | | | | | |
| 3. I'm learning how to improve my professional practice. | | | | | | |
| 4. What I learn is in line with my professional practice. | | | | | | |

☐ Reflexive thinking

| Answers | Not yet answered | Almost never | Rarely | Sometimes | Often | Almost always |
|---|---|---|---|---|---|---|
| In this online course ... | | | | | | |
| 5. I take a critical look at the way I learn. | | | | | | |
| 6. I take a critical look at my own ideas. | | | | | | |
| 7. I take a critical look at the ideas of other students. | | | | | | |
| 8. I take a critical look at the ideas developed in the documents. | | | | | | |

☐ Interactivity

| Answers | Not yet answered | Almost never | Rarely | Sometimes | Often | Almost always |
|---|---|---|---|---|---|---|
| In this online course ... | | | | | | |
| 9. I explain my ideas to other students | | | | | | |
| 10. I ask the other students to explain their ideas. | | | | | | |
| 11. The other students ask me to explain my ideas. | | | | | | |
| 12. The other students respond to my ideas. | | | | | | |

☐ Teacher's Support

| Answers | Not yet answered | Almost never | Rarely | Sometimes | Often | Almost always |
|---|---|---|---|---|---|---|
| In this online course … | | | | | | |
| 13. The teacher stimulates my thinking. | | | | | | |
| 14. The teacher encourages me to participate. | | | | | | |
| 15. The teacher gives the example in expression | | | | | | |
| 16. The teacher sets the example in self-criticism | | | | | | |

☐ Peer assistance

| Answers | Not yet answered | Almost never | Rarely | Sometimes | Often | Almost always |
|---|---|---|---|---|---|---|
| In this online course … | | | | | | |
| 17. The other students encourage my participation. | | | | | | |
| 18. The other students congratulate me on my contribution. | | | | | | |
| 19. The other students appreciate my contribution. | | | | | | |
| 20. The other students take part in my efforts to learn. | | | | | | |

☐ Interpretation

| Answers | Not yet answered | Almost never | Rarely | Sometimes | Often | Almost always |
|---|---|---|---|---|---|---|
| In this online course … | | | | | | |
| 21. I understand the messages of other students | | | | | | |
| 22. Other students understand my messages | | | | | | |
| 23. I understand the teacher's messages | | | | | | |
| 24. The teacher understands my messages | | | | | | |

| 25. How long did it take you to respond to this consultation? | |

| 26. Do you have any other comments? |

▷ **APPENDIX 09: Assessment by learner: a customized survey form**

**The use of this questionnaire is completely anonymous. Your assessment will be useful in improving the quality of future trainings.**

▷ **Did you previously look up to the website dedicated to this training?**

- ☐ Yes
- ☐ No

▷ **Did you already have knowledge of distance learning methods before taking this training?**

- ☐ Yes
- ☐ No

▷ **Was the training appropriate to your social lifestyle outside of the university?**

- ☐ Yes
- ☐ Rather
- ☐ Not really
- ☐ No
- ☐ No answer

▷ **In your opinion, did this distance learning give the same results as in face-to-face?**

- ☐ Absolutely
- ☐ Suitably
- ☐ Not quite
- ☐ Not at all
- ☐ No answer

▷ **How is this distance learning different from your face-to-face learning? (Many possible responses)**

- ☐ More individual autonomy
- ☐ No more risk of loneliness
- ☐ More flexible in working time
- ☐ More collaborative activity
- ☐ A new form of monitoring by ' teaching
- ☐ A new form of ' assessment
- ☐ Other: ……………………………

▷ **What types of constraints (permanent or occasional) did you meet during your remote monitoring of the learning?**

☐ Not enough time to follow all synchronous sessions

☐ Not very powerful computer

☐ Slow internet connection

☐ Lack of support from the tutor

☐ Other: ................................

☐ No answer

▷ **How were your conditions of remote access to the platform for this course (equipment used, Internet connection)?**

☐ Excellent

☐ Good

☐ Fair

☐ Bad

☐ No answer

▷ **Was the duration of the training appropriate?**

☐ Too short

☐ Adapted

☐ Too long

☐ No answer

▷ **Was the time devoted to practical exercises or scenarios during the modules adequate?**

☐ Yes

☐ No

☐ No answer

▷ **Did the proposed teaching methods generally suit you?**

☐ Yes

☐ Rather

☐ Not really

☐ No

☐ No answer

▷ **How do you judge the performance of your tutor?**

  ☐ Excellent
  ☐ Good
  ☐ Fair
  ☐ Bad
  ☐ No answer

▷ **What do you think of the quality of the materials used during the learning? (Presentations, documents, video …)**

  ☐ Excellent
  ☐ Good
  ☐ Fair
  ☐ Bad
  ☐ No answer

▷ **Beyond the content, was this training an enriching experience?**

  ☐ Yes
  ☐ No
  ☐ No answer

▷ **What you appreciated most in this training:**

  ………………………………………………………………………………………………………………………
  ………………………………………………………………………………………………………………………
  ………………………………………………………………………………………………………………………
  ………………………………………………………………………………………………………………………

▷ **What did you like least and what are your proposals to remedy it?**

  ………………………………………………………………………………………………………………………
  ………………………………………………………………………………………………………………………
  ………………………………………………………………………………………………………………………
  ………………………………………………………………………………………………………………………
  ………………………………………………………………………………………………………………………

▷ APPENDIX 10: assessment by the tutor: a customized survey form

The data collected through this form will be used to improve the conditions for monitoring learners during future trainings

▷ **Did you already have knowledge of distance learning methods before this experience?**

- ☐ Yes
- ☐ No

▷ **Have you ever tutored distance learning?**

- ☐ Yes
- ☐ No

▷ **Have you already received training on e-tutoring?**

- ☐ Yes
- ☐ No

▷ **Did you consult the website dedicated to this training?**

- ☐ Yes
- ☐ No
- ☐ No answer

▷ **In your opinion, did this distance learning give the same result as a face-to-face learning?**

- ☐ Absolutely
- ☐ Suitably
- ☐ Not quite
- ☐ Not at all
- ☐ No answer

▷ **How is this distance learning different from your face-to-face practice? (Many possible responses)**

- ☐ More autonomy for the learning
- ☐ No more risk of loneliness
- ☐ More flexible in working time
- ☐ More collaborative activity
- ☐ A new form of monitoring learners
- ☐ A new form of assessment
- ☐ Other: ……………………………….

▷ **What kinds of constraints (permanent or occasional) did you meet during the tutoring\* of distance learners?**

☐ Not enough time to follow the learners individually

☐ Slow internet connection

☐ Little reactivity of learners

☐ Few learners in synchronous sessions

☐ Other: ……………………………………………………….

☐ No answer

▷ **How were your technical conditions to provide this distance tutoring\* (equipment used, Internet connection)?**

☐ Excellent

☐ Good

☐ Fair

☐ Bad

☐ No answer

▷ **Was the length of the training period appropriate?**

☐ Too short

☐ Adapted

☐ Too long

☐ No answer

▷ **Was the time devoted to practical activities suitable for learners?**

☐ Yes

☐ No

☐ No answer

▷ **Do you find suitable and sufficient the learning activities you have tutored?**

☐ Yes

☐ Rather

☐ Not really

☐ No

☐ No answer

▷ **How do you estimate the learners' autonomy during the online course?**

☐ Excellent

- ☐ Brillante
- ☐ Perfect
- ☐ No answer

▷ **What do you think about the quality of learning materials you used? (Presentations, documents, video ...)**

- ☐ Excellent
- ☐ Good
- ☐ Fair
- ☐ Bad
- ☐ No answer

▷ **Was this training an enriching experience for you?**

- ☐ Yes
- ☐ No
- ☐ No answer

▷ **What you appreciated most in your tutoring\* experience?**

...................................................................................................................................................
...................................................................................................................................................
...................................................................................................................................................
...................................................................................................................................................

▷ **What did you like least and what are your proposals to remedy it?**

...................................................................................................................................................
...................................................................................................................................................
...................................................................................................................................................
...................................................................................................................................................

▷ **APPENDIX 11: Record form and project closing report**

The validation of the project requires an acceptable level of operability of the system during the period of implementation of the blended learning. The following indicators are to be completed by the project team and validated by the project manager.

▷ **VALIDATION OF PROJECT PRODUCTS**

| ■ *Technical aspects:* | YES | NO |
|---|---|---|
| ▪ Is the platform hosted on an institution server? | ☐ | ☐ |
| ▪ Can we access the platform from all types of mobile terminals? | ☐ | ☐ |
| ▪ Was the connection often stable during online learning? | ☐ | ☐ |
| ▪ Was the platform often stable to go through with the learning? | ☐ | ☐ |
| ▪ … | ☐ | ☐ |
| ■ *Management and administrative aspects* | | |
| ▪ Does the institution have an internal service or does it have a maintenance contract with a service company or an external expert for the intervention and periodic maintenance of the platform? | ☐ | ☐ |
| ▪ Has the institution appointed a remote learning administrator? | ☐ | ☐ |
| ▪ Has the institution planned an annual budget for online learning? | ☐ | ☐ |
| ▪ Does the project team have well-distributed roles and attributions which cover the different needs of online learning? | ☐ | ☐ |
| ▪ Have all the team members followed the training of trainers planned in step 01 of phase 02 of the project? | ☐ | ☐ |
| ▪ Have all the courses planned for the test phase been finalized and hosted on the platform? | ☐ | ☐ |
| ▪ … | ☐ | ☐ |
| ■ *Educational aspects* | | |
| ▪ Did the learners carry out most of the learning activities? | ☐ | ☐ |
| ▪ Has the dropout rate exceeded half of those enrolled? | ☐ | ☐ |
| ▪ Were the presence statistics on the platform more than the average (use traces history on Moodle)? | ☐ | ☐ |
| • Did the work done by the learners exceeded the average of all planned activities? | ☐ | ☐ |
| • Were the tutors often present in the synchronous sessions? | ☐ | ☐ |
| • … | ☐ | ☐ |
| ■ *Educational aspects* | | |
| • Did the project team assimilate the new learning methods introduced by eLearning? | ☐ | ☐ |
| • According to the results of the learning assessment questionnaire (**Appendices 07, 08**), were the majority of the learners satisfied with the distance learning followed in the test phase? | ☐ | ☐ |
| • According to the results of the tutoring* assessment questionnaire (**Appendix 08**), were the tutors mostly satisfied with the distance learning proposed in the project? | ☐ | ☐ |
| • Were the tutors satisfied with their tutoring* experiences (**Appendix 08**)? | ☐ | ☐ |
| • Is the teaching team persuaded to reproduce this teaching method in other disciplines and departments? | ☐ | ☐ |
| • … | ☐ | ☐ |
| | Result | |

▷ **CLOSING REPORT AND DELIVERY OF THE PROJECT**

Based on the previous results and mutual consensus between the project team and the validation committee, a report is drawn up to validate the functioning of the online learning system and deliver it to [Name of the institution].

In this regard, it is agreed between [Name and institution] and the validation committee, the following:
- ...........................................................................................................................................................................
- ...........................................................................................................................................................................
- ...........................................................................................................................................................................
- ...........................................................................................................................................................................

After validation of the eLearning project by the two parties, [Name of the institution] now takes full responsibility for the system and will ensure its supply, maintenance and operation according to its university educational policy.

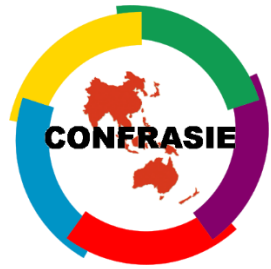

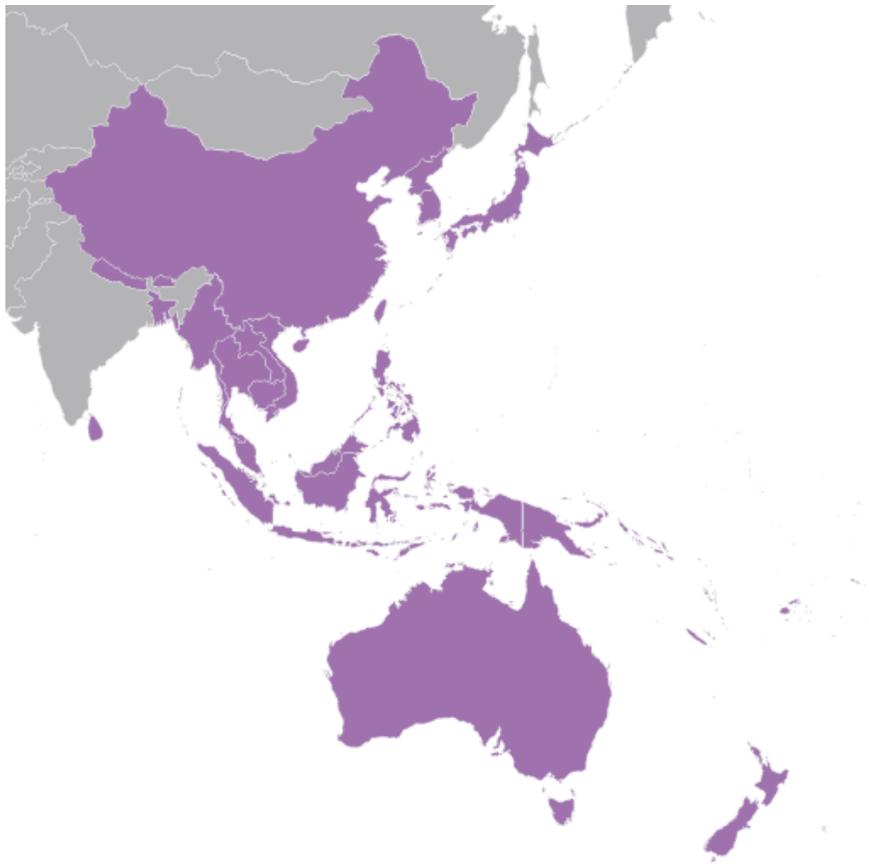

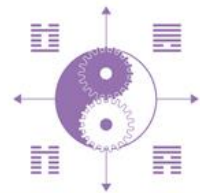

**Direction régionale Asie-Pacifique**
21, Le Thanh Tong – Hoan Kiem – Hanoï – Vietnam
www.auf.org/asie-pacifique